\pdfoutput=1
	\documentclass[aps,prb,onecolumn,superscriptaddress,noshowpacs,longbibliography]{revtex4-2}

	\usepackage{amsmath,amsfonts,amsbsy,amssymb,amsthm}
	\usepackage{comment}
	\usepackage{graphicx}
	\usepackage{esint}
    \usepackage{amsopn}
	\usepackage{tipa}
	\usepackage{braket}
	\usepackage{bm}
	\usepackage{hyperref}
	\hypersetup{
	    unicode=false,          
	    pdftoolbar=false,        
	    pdfmenubar=true,        
	    pdffitwindow=false,     
	    pdfstartview={FitH},    
	    pdfkeywords={MBL} {AGP} {Mass-deformed SYK}, 
	    pdfnewwindow=true,      
	    colorlinks=true,       
	    linkcolor=red,          
	    citecolor=blue,        
	    filecolor=magenta,      
	    urlcolor=blue           
	}
	\usepackage{xcolor}
	\usepackage[page]{appendix}
	\newcommand{\reffig}[1]{\mbox{Fig.~\ref{#1}}}
	\newcommand{\refsec}[1]{\mbox{Sec.~\ref{#1}}}

	\newcommand{\T}{${\mathcal T}\,$}
	\newcommand{\Ti}{${\mathcal T}$}
	
	\usepackage[	  amsmath	]{empheq}
	\usepackage{amsfonts}
	\usepackage{epsfig,float}
	\usepackage{times}
	\usepackage{verbatim}   
	\usepackage{color}      

	\def\ba{\begin{align}}
		\def\ea{\end{align}}
	\def\be{\begin{equation}}
		\def\ee{\end{equation}}
	
	\def\bea{\begin{eqnarray}}
		\def\eea{\end{eqnarray}}

	\newcommand{\roughly}[1]{\mathrel{\raise.3ex\hbox{$#1$\kern-0.85em
				\lower1ex\hbox{$\sim$}}}}

   \newcommand{\pcsadd}{Center for Theoretical Physics of Complex Systems, Institute for Basic Science (IBS), Daejeon, Korea, 34126}
	\newcommand{\ustadd}{Basic Science Program, Korea University of Science and Technology (UST), Daejeon 34113, Republic of Korea}

\usepackage[T1]{fontenc}

\begin{document}
		
\title{Haldane graphene billiards versus relativistic neutrino billiards}
	\author{Dung Xuan Nguyen}
	    \email{dungmuop@gmail.com}
	    \affiliation{\pcsadd}
	    \affiliation{\ustadd}

	\author{Barbara Dietz}
	    \email{bdietzp@gmail.com}
	    \affiliation{\pcsadd}
	    \affiliation{\ustadd}

	\date{\today}

	\begin{abstract}

	We study fluctuation properties in the energy spectra of finite-size honeycomb lattices -- graphene billiards -- subject to the Haldane-model onsite potential and next-nearest neighbor interaction at critical points, referred to as Haldane graphene billiards in the following. The billiards had the shapes of a rectangular billiard with integrable dynamics, one with chaotic dynamics, and one whose shape has, in addition, threefold rotational symmetry. It had been shown that the spectral properties of the graphene billiards coincide with those of the nonrelativistic quantum billiard with the corresponding shape, both at the band edges and in the region of low energy excitations around the Dirac points at zero energy. There, the dispersion relation is linear and, accordingly, the spectrum is described by the same relativistic Dirac equation for massless spin-1/2 particles as relativistic neutrino billiards, whose spectral properties agree with those of nonrelativistic quantum billiards with violated time-reversal invariance. Deviations from the expected behavior are attributed to differing boundary conditions and backscattering at the boundary, which leads to a mixing of valley states corresponding to the two Dirac points, that are mapped into each other through time reversal. We employ a Haldane model to introduce a gap at one of the two Dirac points so that backscattering is suppressed in the energy region of the gap and demonstrate that there the correlations in the spectra comply with those of the neutrino billiard of the corresponding shape.   

	\end{abstract}
	\bigskip
	\maketitle

	\section{Introduction\label{Intro}}
	Due to the extraordinary band structure of graphene, a crystalline monolayer of carbon atoms arranged on a honeycomb lattice~\cite{Wallace1947,Boehm1994}, that entail relativistic phenomena~\cite{Beenakker2008,Neto2009}, its pioneering fabrication~\cite{Novoselov2004} induced numerous experimental and theoretical investigations. Namely, the conduction and valence bands touch each other conically at the Fermi energy~\cite{Beenakker2008,Neto2009}, implicating a linear dispersion relation, so  that in these regions the electronic properties of graphene are described by the Dirac equation for massless Dirac fermions~\cite{Semenoff1984,Katsnelson2007}. Thus, even though the Fermi velocity of the electrons is by a factor of 300 smaller than that of light, it features relativistic phenomena~\cite{Geim2007,Avouris2007,Beenakker2008,Neto2009,Abergel2010} like pseudodiffusive transport~\cite{Zhang2008,Zandbergen2010,Bittner2012}, the quantum Hall effect~\cite{Poo2011}, Zitterbewegung~\cite{Zhang2008}, and edge states~\cite{Zandbergen2010,Kuhl2010}. Accordingly, the touch points are commonly referred to as Dirac points (DPs).

	The band structure of graphene originates from the hexagonal lattice structure which is formed by two independent, interpenetrating triangular lattices with threefold rotational ($C_3$) symmetry. The Dirac points, $K$ and $K^\prime$, associated with the two triangular lattices, are at the corners of the Brillouin zone (BZ). Because of the relativistic features exhibited by graphene in the regions around the DPs, these are referred to as \emph{relativistic} regions~\cite{Dietz2013,Neto2009}. At the center of the Brillouin zone, the $\Gamma$ point, the conduction band has a maximum and the valence band has a minimum and their shapes are parabolic implying that the Dirac equation is no longer applicable. Accordingly, the band structure can be divided into \emph{nonrelativistic} regions around the $\Gamma$ points~\cite{Dietz2013} and \emph{relativistic} ones around the $K$ and $K^\prime$ points. At its saddle points a topologcal transition from the conically to the parabolically shaped band structure, that is from the relativistic to the nonrelativistic regions takes place~\cite{Dietz2013}. There the group velocity vanishes for an infinitely extended graphene sheet and the density of states (DOS) exhibits a logarithmic divergence~\cite{Neto2009}. These are Van Hove singularities that generally occur in two-dimensional crystals with a periodic structure~\cite{VanHove1953}.

	The linear dispersion relation of graphene arises from the symmetry properties of its honeycomb structure~\cite{Slonczewski1958}, particularly time-reversal (\Ti) symmetry, inversion symmetry, and $C_3$ symmetry. Thus, any system subject to a spatially periodic potential with hexagonal structure like, e.g., a photonic crystal~\cite{Parimi2004,Joannopoulos2008}, may comprise energy regions, where they are effectively described by the relativistic Dirac equation for spin-$1/2$ particles. Indeed, numerous realizations of artificial graphene~\cite{Polini2013} popped up soon after the fabrication of gaphene. Examples are two-dimensional electron gases subject to a potential on a honeycomb lattice~\cite{Singha2011,Nadvornik2012}, molecular assemblies arranged on a copper surface~\cite{Gomes2012}, ultracold atoms in optical lattices~\cite{Tarruell2012,Uehlinger2013} and photonic crystals~\cite{Bittner2010,Kuhl2010,Sadurni2010,Bittner2012,Bellec2013,Bellec2013a,Rechtsman2013,Rechtsman2013a,Khanikaev2013,Wang2014,Shi2015}.

In Refs.~\cite{Dietz2013,Dietz2015,Dietz2016,Zhang2023} several thousands of eigenfrequencies of superconducting microwave photonic crystals, so-called 'Dirac microwave billiards', were determined experimentally. Such devices emulate finite-size artificial graphene billiards (GBs) introduced in~\cite{Libisch2009,Wurm2009} as a model for graphene quantum dots~\cite{Miao2007,Ponomarenko2008,Westervelt2008,Schnez2009,Guettinger2010,Guettinger2012}, that became of interest because they exhibit non-relativistic \emph{and} relativistic features~\cite{DiVincenzo1984,Novoselov2004,Geim2007,Avouris2007,Miao2007,Ponomarenko2008,Beenakker2008,Zhang2008,Neto2009,Abergel2010,Zandbergen2010}. Graphene billiards are constructed by cutting out of an extended honeycomb lattice a sheet with corresponding shape and their eigenstates are computed based on a tight-binding model (TBM) with Dirichlet BCs on the next-nearest outer sites along the boundary. Their properties have been shown to agree with those of Dirac microwave billiards in~\cite{Dietz2013,Dietz2015,Dietz2016,Zhang2023} when taking into account up to third-nearest neighbor hopping. Actually to be more precisely, Dirac billiards exhibit two Dirac points and emulate the properties of a combination of a honeycomb and kagome lattice~\cite{Maimaiti2020,Zhang2021}. Yet, it was demonstrated in these works and in Refs.~\cite{Dietz2015,Dietz2016,Zhang2023} that around the lower DP the properties of their eigenstates are well described by those of the GB of corresponding shape. 
	
The microwave Dirac billiards considered in~\cite{Dietz2013,Dietz2015,Dietz2016,Zhang2023} had the shapes of an integrable rectangular billiard, a chaotic Africa billiard and a chaotic one with a $C_3$ symmetry. The objective of the experiments was to investigate in the context of quantum chaos the spectral properties of GBs in the relativistic region, for which complete and long eigenvalue sequences are needed. A classical billiard (CB) is a bounded two-dimensional domain, in which a point particle moves freely and is reflected specularly at the boundary. Since the classical dynamics of billiards only depends on the shape of their domain~\cite{Sinai1970,Bunimovich1979,Berry1981}, they provide a paradigm model for the search of signatures of classical chaos in the corresponding quantum system, which is the primary objective of quantum chaos. The eigenstates of the corresponding nonrelativistic quantum billiard (QB) are obtained from the solutions of the Schr\"odinger equation for a free particle by imposing Dirichlet boundary conditions (BCs) on the wave functions. Berry and Mondragon proposed relativistic neutrino billiards~\cite{Berry1987} (NBs). Their spinor eigenstates are solutions of the Dirac equation for a massless spin-1/2 particle with the BC that there is no outgoing flow. 

The central question of the studies with such billiard systems was, whether the spectral properties comply with those of typical quantum systems with integrable or chaotic dynamics. One key aspect of quantum chaos are the fluctuation properties in the eigenvalue spectrum of a quantum system, and their connection to the properties of the dynamics of the corresponding classical system. Berry and Tabor showed in~\cite{Berry1977a} that the eigenvalues of typical integrable systems~\cite{Robnik1998} exhibit the same fluctuation properties as Poissonian random numbers. It was speculated in Refs.~\cite{Berry1979,Casati1980} and then stated in a conjecture by Bohigas, Giannoni and Schmit~\cite{Bohigas1984} that the spectral properties of typical quantum systems with a chaotic classical dynamics are well described by those of random matrices from the Gaussian ensembles of corresponding universality class~\cite{Mehta1990,LesHouches1989,Guhr1998,Haake2018}, namely the Gaussian orthogonal (GOE) and unitary (GUE) ensemble, when \T invariance is preserved and violated, respectively. The Dirac Hamiltonian associated with NBs is not \T invariant. Therefore, the spectral properties of typical NBs with the shapes of chaotic CBs and no geometric symmetry agree well with those of random matrices from the GUE~\cite{Berry1987}. Since GBs are governed in the conical valley regions around the Dirac points by this Dirac Hamiltonian, their spectral properties were expected to exhibit similar features as the NB of corresponding shape, that is, GUE statistics if the shape is that of a chaotic CB. This assumption was confirmed in experiments with graphene quantum dots~\cite{Guettinger2008,Guettinger2010,Ponomarenko2008}. However, numerical studies~\cite{Libisch2009,Wurm2009} and the experimental investigations with superconducting microwave Dirac billiards~\cite{Dietz2013,Dietz2015,Dietz2016,Zhang2023} revealed that they are conform with those of nonrelativistic QBs of corresponding shape, that is, with GOE statistics. 
	
The discrepancies were attributed to the BCs~\cite{Rycerz2012,Rycerz2013} and to back scattering at the boundary, which leads to a mixing of valley states around the $K$ and $K^\prime$ points~\cite{Wurm2009}. To be more explicitly, \T invariance is violated in each solitary Dirac cone, where the electronic excitations are effectively described by the same relativistic Weyl equation~\cite{Weyl1929}, also referred to as two-dimensional Dirac equation~\cite{Beenakker2008}, for spin-1/2 particles as in a NB. However, it is restored in GBs due to the occurrence of two independent Dirac cones that are mapped into each other when applying the associated time-reversal operator $\hat T$. Therefore, agreement with GOE statistics is expected for GBs with a chaotic shape, because the back scattering at the boundary induces an intervalley scattering~\cite{Rycerz2012}. If this indeed is the reason, then the properties should coincide with those of a relativisctic NB, if the scattering from one Dirac cone to the other one is prevented. The objective of the present work is to demonstrate that the spectral properties of GBs coincide with those of the corresponding NB, when introducing a gap at the $K^\prime$ point, implicating that within the energy range of the gap the eigenstates are confined to the conical valley around the $K$ point. This is achieved based on the Haldane model~\cite{Haldane1988} on a honeycomb lattice, which has the particular property that nonzero quantization of the Hall conductance occurs even though no external magnetic field is applied. Indeed, similar to NB, the Haldane model explicitly breaks time-reversal symmetry due to a purely imaginary next-to-nearest neighbor hopping term. In~\refsec{HM} we briefly introduce the honeycomb-lattice based Haldane model. To realize a GB exhibiting near the Fermi energy the properties of a relativistic NB of corresponding shape, we employ a finite-size version of the Haldane lattice structure, named Haldane GBs in the following. In~\refsec{NR} we present numerical results for Haldane GBs with the shape of a rectangular, a Africa and a $C_3$-symmetric CB. Finally, in~\refsec{Concl} we discuss our results and comment on a possible experimental realization of the Haldane model.    

\section{Haldane model\label{HM}}
	The Haldane model was originally introduced by Duncan Haldane in \cite{Haldane1988} based on the TBM for graphene. It provides the simplest 2D model that acquires the Quantum Hall effect (QHE), despite absence of an external magnetic field and the associated Landau levels. Kane and Mele \cite{KaneMele2005} generalized it and thereby developed the simplest model for a topological insulator exhibiting the spin Hall effect by doubling the Haldane model. Interestingly, for a particular choice of the Haldane mass $M$ and the purely imaginary next-to-nearest hopping parameter $\tilde t_2=it_2$ with $t_2\in\mathbb{R}$, the band structure generated within the Haldane model shows only a single Dirac cone at the $K$ or the $K'$ point and is gapped at the other one. Accordingly, we expected that for such critical values within the energy range of the gap the spectral properties of a Haldane GB are similar to those of the NB~\cite{Berry1987} of corresponding shape. We will illustrate for three geometries of Haldane GBs that this indeed is the case. This section summarizes the tight-binding construction of the Haldane model for self-contained purposes.
	
		\subsection{Tight-binding model of graphene}
\begin{figure}
	\centering
	\begin{tabular}{@{}c@{}}
		\begin{tabular}{@{}c@{}}
			\includegraphics[width=.4\linewidth]{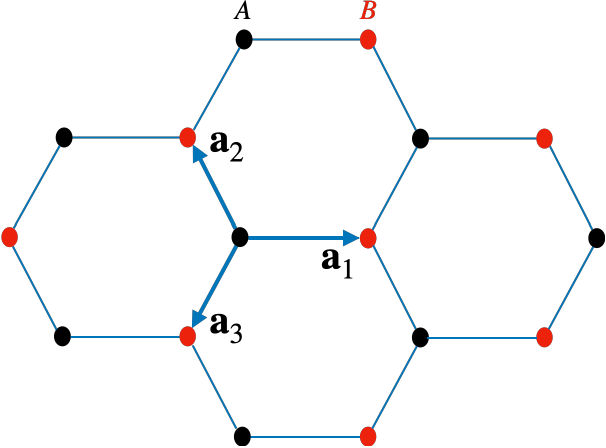} \\[\abovecaptionskip]
			\small (a) 
	\end{tabular}

	\begin{tabular}{@{}c@{}}
			\includegraphics[width=.3\linewidth]{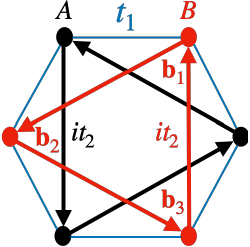} \\[\abovecaptionskip]
			\small (b) 
		\end{tabular}
	\end{tabular}	
	\caption{(a) The honeycomb structure of graphene. (b) The nearest neighbor hopping term $t_1$ and the purely imaginary next-to-nearest neighbor Haldane tunneling term $i t_2$. }\label{fig:TB}
\end{figure}

	We begin with the TBM of graphene~\cite{Neto2009}, illustrated in~\reffig{fig:TB}(a), which is formed by two triangular sublattices $A$ and $B$. The vectors $\mathbf{a}_i$ are defined as
$
	    \mathbf{a}_1=(a,0),  \mathbf{a}_2=(-\frac{a}{2}, \frac{\sqrt{3}a}{2}),  \mathbf{a}_3=(-\frac{a}{2}, -\frac{\sqrt{3}a}{2}) 
$,
	where $a$ is the distance between neighboring sites of the honeycomb lattice. In the numerical simulations we set it to unity, $a=1$. 
	The TBM Hamiltonian of graphene is given by~\cite{Neto2009} 
	\begin{equation}
	    \hat H^{0}= t_1\sum_{\langle i,j \rangle} \left( \hat{a}^\dagger_i \hat{b}_j + h.c \right). 
	\end{equation}
	Here, $\hat{a}_i$ and $\hat{b}_j$ denote the operators that annihilate an electron at sites $A_i$  and $B_j$, respectively, and the notation $\langle i,j \rangle$ indicates that summations are performed over nearest-neighbor sites. In the momentum space, the Hamiltonian in the basis 
	$
	\psi^T=\left(\psi_A(\mathbf{k}),\psi_B(\mathbf{k}) \right)
	$ becomes~\cite{Dresselhaus1996}
	\begin{equation}
	    \hat H^0(k) = t_1 {\sum}_{i=1}^3  \big( \hat\sigma_x \cos( \mathbf{k} \cdot \mathbf{a}_i)  - \hat\sigma_y \sin(\mathbf{k} \cdot \mathbf{a}_i) \big) ,
	\end{equation}
	where $\hat\sigma_x,\hat\sigma_y,\hat\sigma_z$ denote the Pauli matrices. The associated dispersion exhibits a particle-hole symmetry \cite{Neto2009}.
	Near the Dirac cone $\mathbf{K}=\left(\frac{2\pi}{3 a},\frac{2\pi}{3 \sqrt{3}a}\right)$ in the momentum space, the effective wave equation of electrons is given in terms of the Dirac Hamiltonian for a spin-1/2 particle,
	\begin{equation}
	   \hat H^0_K(\mathbf{q})=\frac{3 t_1 a}{2}\boldsymbol{\hat\sigma}\cdot\mathbf{q}, \qquad (\mathbf{k}= \mathbf{K}+ \mathbf{q})
	\end{equation}
with $\boldsymbol{\hat\sigma}=({\hat\sigma}_x,{\hat\sigma}_y)$.
Similarly, near the Dirac cone $\mathbf{K}'=-\mathbf{K}$ it is given by
	\begin{equation}
	    \hat H^0_{K'}(\mathbf{q})=\frac{3 t_1 a}{2} \boldsymbol{\hat\sigma^\ast}\cdot\mathbf{q}, \qquad (\mathbf{k}= \mathbf{K}'+ \mathbf{q}),
	\end{equation}
	where $\boldsymbol{\hat\sigma^\ast}$ denotes complex conjugation of $\boldsymbol{\hat\sigma}$.
	The appearance of the two Dirac cones is protected by time-reversal symmetry, inversion symmetry, and the discrete rotational symmetry $C_3$ of the honeycomb-lattice structure~\cite{Beenakker2008,Bernevigbook2013}. 
	\subsection{Haldane tunneling and mass terms}
	The extension to the Haldane model is illustrated in~\reffig{fig:TB}(b). Here, the next-to-nearest neighbor site vectors  are 
$
	     \mathbf{b}_1=(0,\sqrt{3} a),   \mathbf{b}_2=(-\frac{3 a}{2}, -\frac{\sqrt{3}a}{2}),  \mathbf{b}_3=(\frac{3 a}{2}, -\frac{\sqrt{3}a}{2}).
	$
	Following Haldane, we introduce, in addition to the nearest-neighbor hopping, a nonzero purely imaginary next-to-nearest neighbor hopping parameter, $i t_2$~\cite{Haldane1988}. Furthermore, we introduce onsite potentials $M$ with $M>0$ on all sites of sublattice $A$ and $-M$ on all sites of sublattice $B$. In momentum space the resulting Hamiltonian is given by 
	\begin{equation}
	\label{eq:HamHD}
		\hat H(\mathbf{k})=\hat H^0(\mathbf{k})+ \left(M- 2 t_2 \sum_{j=1}^3 \sin (\mathbf{k}\cdot \mathbf{b}_j)  \right)\hat\sigma_z
	\end{equation}
	The first term in the rounded brackets is the inversion-breaking mass term and the second one induces time-reversal invariance violation. The effective Hamiltonian near the $K$ and the $K'$ points, respectively, becomes 
	\begin{align}
		\hat H_K(\mathbf{q})  &=\frac{3 t_1 a}{2}\boldsymbol{\hat\sigma}\cdot\mathbf{q} + \left(M- 3 \sqrt{3}t_2\right)\hat\sigma_z \\
		\hat H_{K'}(\mathbf{q})  &=\frac{3 t_1 a}{2}\boldsymbol{\hat\sigma^\ast}\cdot\mathbf{q} + \left(M+ 3 \sqrt{3}t_2\right)\hat\sigma_z 
	\end{align}
In the trivial insulator phase of the Haldane model, $|t_2| < \frac{M}{3\sqrt{3}}$, both Dirac cones are gapped. For $|t_2| > \frac{M}{3\sqrt{3}}$, the Haldane model is in a non-trivial topological phase -- the Chern insulator phase -- where both Dirac cones are gapped, but the Chern number is non-zero. Yet, at the critical point $t_2=\frac{M}{3\sqrt{3}}$, one Dirac cone, namely that at the $K'$ point is gapped out with the effective mass $2M$ and that at the $K$ point survives. Accordingly, in the low energy limit $|E|< 2M$, there is only one Dirac cone at the $K'$ point. Vice versa, when $t_2=-\frac{M}{3\sqrt{3}}$, the Dirac cone at the $K$ point is gapped out and there is a single Dirac cone at the $K'$ point. Thus, we expect that at the critical points $t_2=\pm \frac{M}{3\sqrt{3}}$ the fluctuation properties in the energy spectrum of a Haldane GB, which is obtained by cutting out of the honeycomb lattice a sheet with the shape of a certain GB, coincide in the energy window $|E|\leq{\rm min}(\vert t_1\vert/2, 2M)$ with those of the corresponding neutrino billiard. In the numerical simulations, we will study the Haldane model at the critical point $t_2=\frac{M}{3\sqrt{3}}$, and we set $t_1=1$  and $0<M<1/2$, so that ${\rm min}(\vert t_1\vert/2, 2M)=2M$.   

\begin{figure}
\centering
	\includegraphics[width=0.49\linewidth]{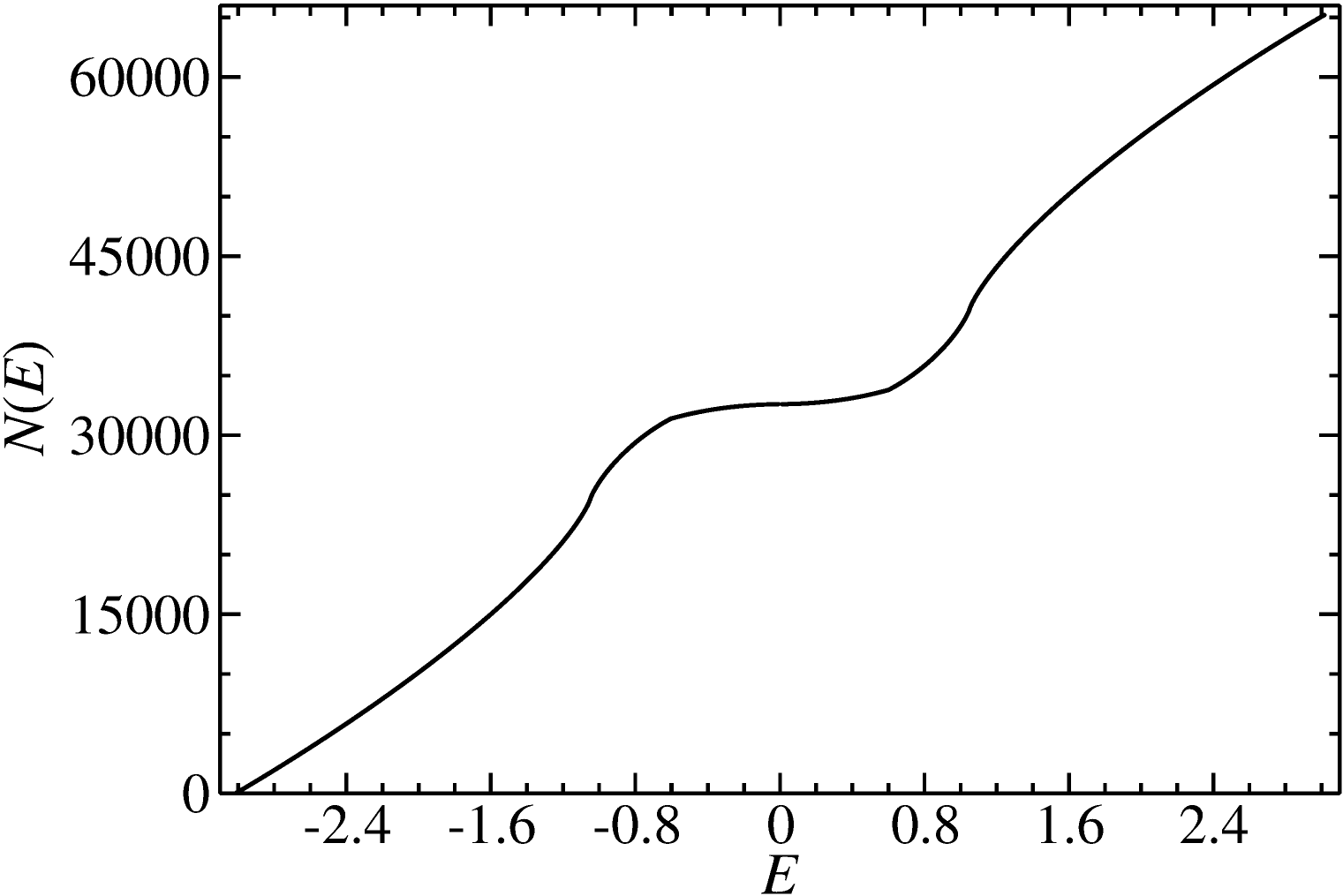}
\includegraphics[width=0.49\linewidth]{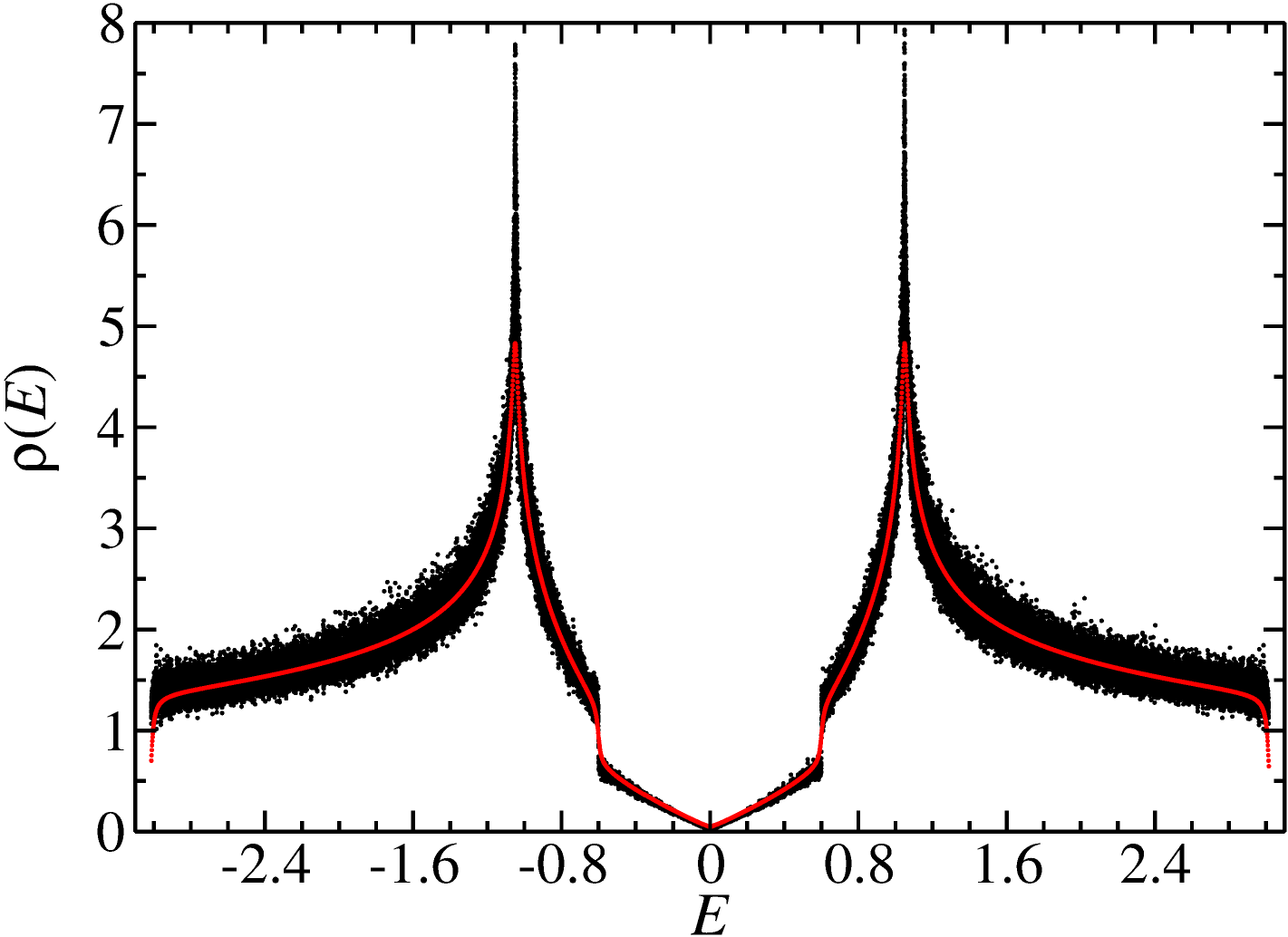}
	\caption{Integrated spectral density $N(E)$ (left) and density of states $\rho(E)$ (right) of the Haldane GB with mass $M=0.3$ and the shape of an Africa billiard. Its lattice consists of $N = 65199$ sites. The DOS exhibits a jump at $E=\pm 2M$, that is at the value where the conduction and valence bands exhibit a minimum and maximum, respectively, at the $K^\prime$ point (see~\reffig{fig:Bandst}).}
    \label{GBDOS}
\end{figure}

\begin{figure}
\centering
	\begin{tabular}{@{}c@{}}
		\begin{tabular}{@{}c@{}}
			\includegraphics[width=.5\linewidth]{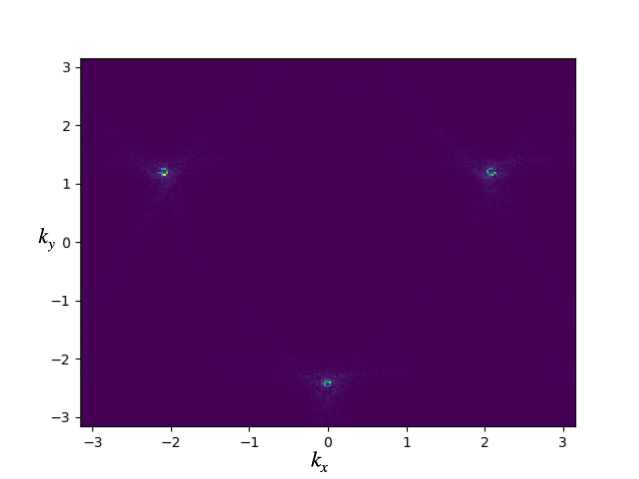} \\[\abovecaptionskip]
			\small (a) 
	\end{tabular}
    \begin{tabular}{@{}c@{}}
			\includegraphics[width=.5\linewidth]{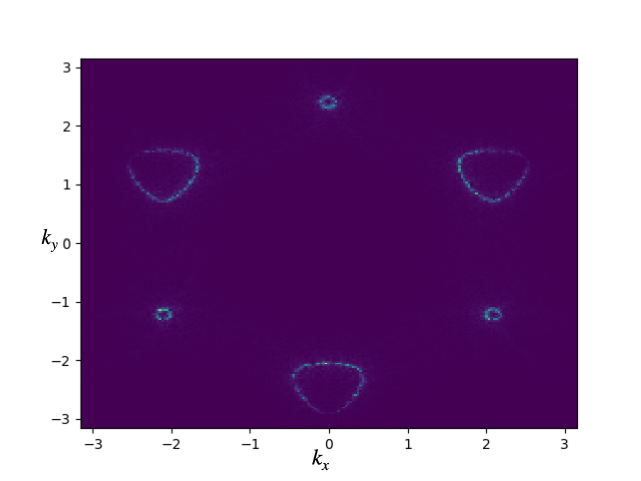} \\[\abovecaptionskip]
			\small (b) 
		\end{tabular}
	\end{tabular}	
	\caption{(a) Momentum distribution of an eigenstate near $E \sim 0$ for the $C_3$ geometry with $M=0.3$. Here we chose one that is invariant under rotation by $\frac{2\pi}{3}$. It is nonvanishing  only in the $K$ valley. (b) Same as (a) for an eigenstate near $E\gtrsim 2M$. Contribution come from the $K$ and $K^\prime$ valleys as expected from the energy spectrum of the critical Haldane model. }\label{fig:WFF}
\end{figure}

\begin{figure}
	\begin{center}
	\includegraphics[width=0.6\textwidth]{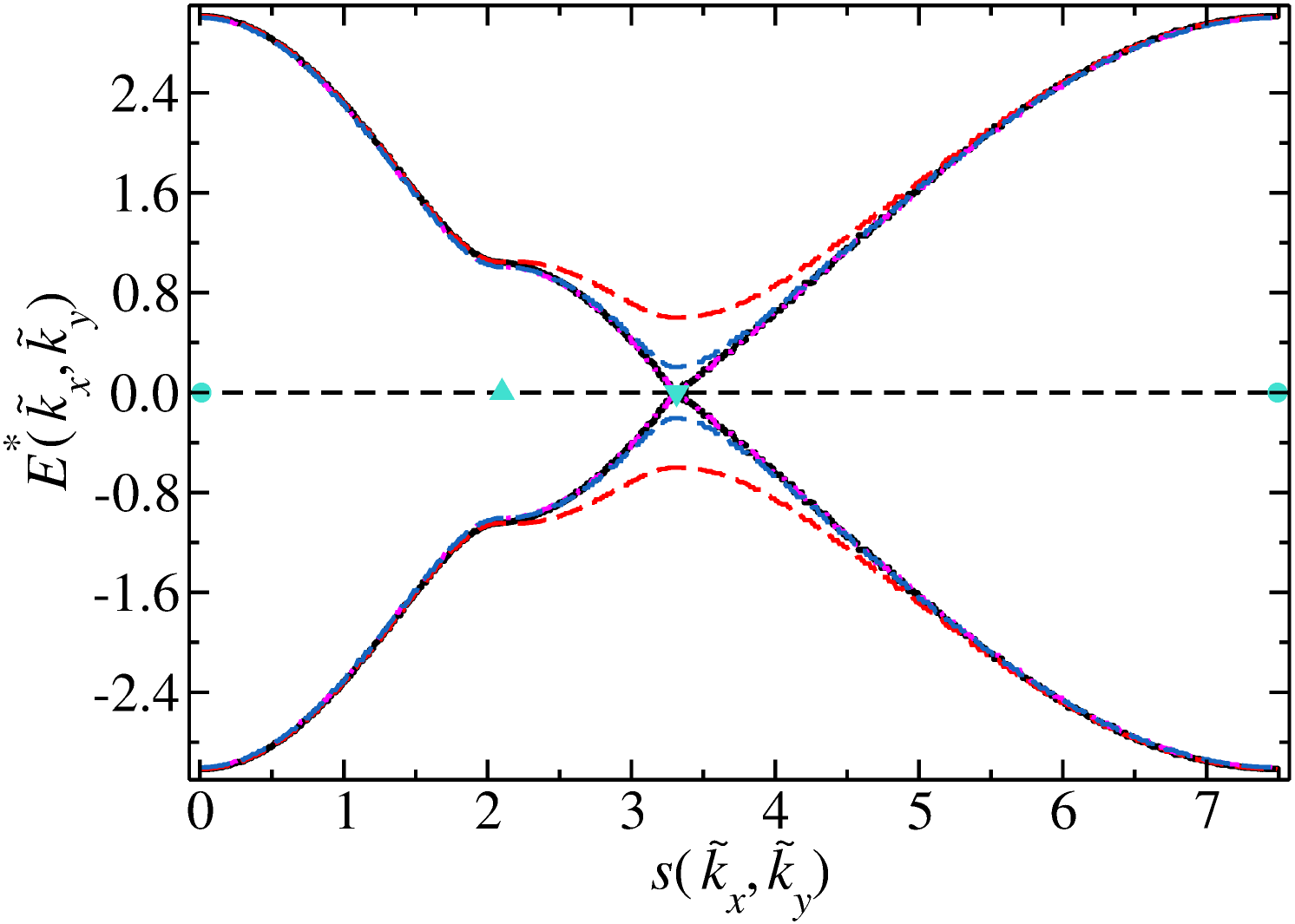}
		\caption{Band structure of the Africa GB. We obtained it by computing for each eigenstate of the GB the momentum distribution $\mathcal{M}(k_x,k_y)$, where we chose the quasimomentum values $(\tilde k_x,\tilde k_y)$ along the path starting from the $\Gamma$ point (turquoise dots) via the saddle point (turquoise triangle) and Dirac point (turquoise down triangle) $K$ (black curve), respectively $K^\prime$ (red dashed curve: $M=0.3$, blue dashed curve: $M=0.1$) back to the $\Gamma$ point. Then we determined for each of the quasimomentum vectors the eigenstate, i.e.,  $n$, for which $\mathcal{M}_n(\tilde k_x,\tilde k_y)$ is maximal. Shown is the associated eigenvalue $E_n=E^\ast(\tilde k_x,\tilde k_y)$ versus the length $s(\tilde k_x,\tilde k_y)$ of the traversed path.}
	\label{fig:Bandst}
\end{center}
\end{figure}

	The eigenstates of a Haldane GB with $N$ sites are obtained by diagonalizing the $N\times N$-dimensional TBM Hamiltonian in configuration space. The integrated spectral density $N(E)=\sum_{n=1}^N\theta(E-E_n)$ with $\theta$ denoting the Heavyside step function, that is, the number of ordered eigenenergies $E_n$, $E_1\leq E_2\leq\dots\leq E_N$, below $E$ is shown in~\reffig{GBDOS} together with the density of states (DOS), $\rho(E)=\frac{\pi^2}{N}\frac{dN(E)}{dE}=\frac{\pi^2}{N}\sum_{n=1}^N\delta(E-E_n)$, for the Haldane GB with the shape of an Africa billiard~\cite{Berry1986} (black line). The red solid line shows the smoothed DOS, which is obtained by replacing the $\delta$ functions by Lorentzians of finite width $\Gamma_L$, 
	\begin{equation}
	\rho^{smooth}(E)=\frac{\pi}{N}\sum_n\frac{\Gamma_{\rm L}}{(E-E_n)^2+\Gamma_{\rm L}^2}, 
	\label{RhoSm}
	\end{equation}
where we chose $\Gamma_{\rm L}=0.01$.  The DOS exhibits a jump at $E=\pm 2M$, that is, at the edges of the band gap appearing at the $K^\prime$ point. The band structure is shown in~\reffig{fig:Bandst}. It was determined based on the momentum distributions, $\mathcal{M}_n(k_x,k_y)$, which is the Fourier transforms of the eigenfunctions $\Psi_n(x,y)$ associated with the eigenvalues $E_n$, from configurational space $(x,y)$ to quasimomentum space $(k_x,k_y)$~\cite{Dietz2015}, 
\begin{equation}
	\mathcal{M}_n(k_x,k_y)=\int_\Omega\Psi_n(x,y)e^{-i(k_x x +k_y y)}dxdy,
\end{equation}
where $\Omega$ is the billiard domain. It exhibits maxima at the wave vector $\boldsymbol{k}=\boldsymbol{k_n}$ corresponding to the eigenvalue $E_n$~\cite{Dietz2015}. In~\reffig{fig:WFF} we show examples for momentum distributions of the Haldane GB with $M=0.3$ $C_3$ symmetry for eigenstates in the relativistic region, one close to the Dirac point and another one close to, but outside the gap region, $E\gtrsim 0.6$. Here we chose eigenstates that are invariant under rotation by $\frac{2\pi}{3}$. Around the Dirac point  $\mathcal{M}_n(k_x,k_y)$ is nonvanishing only at the $K$ points, whereas outside the gap region it is also nonvanishing in the region of the $K^\prime$ point. In~\reffig{fig:Bandst} is plotted the band structure of a Haldane GB with the shape of an Africa billiard and mass $M=0.1$ (blue dashed line) and $M=0.3$ (red dashed line). Here, we chose paths $\boldsymbol{\tilde k}=(\tilde k_x,\tilde k_y)$ in the quasimomentum plane starting at the $\Gamma$ point at $(\tilde k_x,\tilde k_y)=(0,0)$ and continuing via the saddle point at $\boldsymbol{M}=\left(\frac{2\pi}{3a},0\right)$ and the $K$ point (black curve), respectively, the $K^\prime$ point (red and blue curves) back to the $\Gamma$ point and computed $\mathcal{M}_n(\tilde k_x,\tilde k_y)$ for each eigenstate. Plotted is the energy value $E^\ast$ of the eigenstate corresponding to the maximal momentum distribution at $\boldsymbol{\tilde k}$ versus the length of the traversed path.  

\section{Numerical Methods~\label{NM}}
We performed numerical simulations with Haldane GBs with the shapes of a rectangular, Africa and $C_3$-symmetric billiard for masses $0\leq M\leq 0.4$. The results are summarized in~\refsec{NR}. We analyzed their spectral properties and compared them to those of nonrelativistic quantum billiards (QBs) and relativistic neutrino billiards (NBs). The domain $\Omega$ of the billiard is defined in a cartesian coordinate system in polar coordinates, $\boldsymbol{r}=[x(r,\varphi),y(r,\varphi)]$, or in the complex plane, $w(r,\varphi)=x(r,\varphi)+iy(r,\varphi)$, with $\varphi\in [0,2\pi),\, r\in [0,r_0]$, where  the boundary $\partial\Omega$ is obtained by setting $r=r_0$.

The wave equation of nonrelativistic QBs is given by the Schr\"odinger equation of a free particle with Dirichlet BCs along $\partial\Omega$,
\bea
\hat H\psi_m(r,\varphi)&&=-\Delta_{(r,\varphi)}\psi_m(r,\varphi)=k_m^2\psi_m(r,\varphi),\,
\label{Schr}\\
\psi_m(r,\varphi)\vert_{r=r_0}&&=0.\nonumber
\eea
Here, $\psi_m(r,\varphi)$ and $k_m$ denote the eigenfunctions and wavenumbers associated with the eigenvalues $E_m=k_m^2$ of the Hamiltonian $\hat H$. Neutrino billiards~\cite{Berry1987} are governed by the Weyl equation~\cite{Weyl1929} for a non-interacting spin-1/2 particle of mass $m_0$, commonly referred to as Dirac equation in that context,
\be
\boldsymbol{\hat H}_D\boldsymbol{\psi}=\left[c\boldsymbol{\hat\sigma}\cdot\boldsymbol{\hat p}+m_0c^2\hat\sigma_z\right]\boldsymbol{\psi}
=E\boldsymbol{\psi},\, \boldsymbol{\psi}=
\begin{pmatrix}
\psi_1 \\ \psi_2
\end{pmatrix},
\label{DE}
\ee
with the BC
\be
\psi_2(\varphi)=i e^{i\alpha(\varphi)}\psi_1(\varphi).
\label{BC1}
\ee
Here, $\boldsymbol{\hat p}=-i\hbar\boldsymbol{\nabla}$ is the momentum of the particle, $\boldsymbol{\hat H}_D$ the Dirac Hamiltionian and  $E=\hbar ck_E=\hbar ck\sqrt{1+\beta^2}$ with $k$ denoting the free-space wave vector and $\beta=\frac{m_0c}{\hbar k}$ is the ratio of the rest-energy momentum and free-space momentum. Furthermore, $\alpha(\varphi)$ is the angle of the outward-pointing normal vector $\boldsymbol{n}=[\cos\alpha(\varphi),\sin\alpha(\varphi)]$ at $w(r_0,\varphi)$ with respect to the $x$ axis. The BC arises from the requirement that the normal component of the local current, that is, of the expectation value of the current operator $\boldsymbol{\hat u}=\boldsymbol{\nabla}_{\boldsymbol{p}}\boldsymbol{\hat H}_D=c\boldsymbol{\hat\sigma}$, 
\be
\boldsymbol{n}\cdot\boldsymbol{u}(\boldsymbol{r})=c\boldsymbol{n}\cdot\left[\boldsymbol{\psi}^\dagger\boldsymbol{\hat\sigma}\boldsymbol{\psi}\right],\label{eq:curr}
\ee
is zero along the boundary. In Ref.~\cite{Berry1987} only the ultrarelativistic case $m_0=0$ was considered. In Refs.~\cite{Dietz2020,Zhang2021} the transition to the nonrelativistic limit was analyzed, which is reached for $E\simeq m_0c^2$~\cite{Baym2018}, that is, for $\beta\gg 1$. 

We computed the eigenvalues and eigenfunctions of the QB and NB by employing the corresponding boundary-integral equation (BIE), which for the QB is given by~\cite{Baecker2003}
\be
u(\varphi^\prime)=\int_0^{2\pi}d\varphi\vert w^\prime(\varphi)\vert Q^{QB}(k;\varphi,\varphi^\prime)u(\varphi),\label{BIEQB}
\ee
with
\be
Q^{QB}(k;\varphi,\varphi^\prime)=i\frac{k}{2}\cos\left[\alpha(\varphi^\prime)-\xi(\varphi,\varphi^\prime)\right]H_1^{(1)}(k\rho).
\label{QQB}
\ee
Here, we introduced the notations $w^\prime(\varphi)=\frac{dw(\varphi)}{d\varphi}$, $\mathcal{L}$ for the perimeter, and $\xi$ and $\rho$ for the phase and modulus of the distance vector $\boldsymbol{r}(\varphi,\varphi^\prime)$ between two points along the boundary, given in the complex plane parametrization as
\be
e^{i\xi(\varphi,\varphi^\prime)}=\frac{w(\varphi)-w(\varphi^\prime)}{\vert w(\varphi)-w(\varphi^\prime)\vert},\, \rho(\varphi,\varphi^\prime)=\vert w(\varphi)-w(\varphi^\prime)\vert .
\ee
Furthermore, $H_{m}^{(1)}(x)$ is the Hankel function of the first kind of order $m$ \cite{Arfken1995} and $u(s)=\partial_n\psi(s)$ refers to the normal derivative of the wave function $\psi(s)$, with $s$ the arc-length parameter,
\be
s(\varphi)=\int_0^\varphi\vert w^\prime(\tilde\varphi)\vert d\tilde\varphi,\, s\in\left[0,\mathcal{L}\right], ds=\vert w^\prime(\varphi)\vert d\varphi .
\ee

The BIE for the first spinor-eigenfunction of the NB is given by~\cite{Dietz2020,Dietz2022} 
\bea
\label{QNB}
&\left(1-\sin\theta_\beta\right)\psi_1^\ast(\varphi^\prime)=\frac{ik}{4}\fint_0^{2\pi}d\varphi\vert w^\prime(\varphi)\vert e^{i\frac{\Delta\Phi(\varphi,\varphi^\prime)}{2}} Q_1^{NB}(k;\varphi,\varphi^\prime)\psi_1^\ast(\varphi),\\
        &Q_1^{NB}(k;\varphi,\varphi^\prime)=\label{QNB1}
	\cos\theta_\beta\left[e^{i\left(\alpha(\phi^\prime)-\alpha(\phi)\right)}-1\right]H_0^{(1)}(k\rho)\\
        +&\left\{\left[1-\sin\theta_\beta\right]e^{i\left(\xi(\phi,\phi^\prime)-\alpha(\phi)\right)}\nonumber
        +\left[1+\sin\theta_\beta\right]e^{-i\left(\xi(\phi,\phi^\prime)-\alpha(\phi^\prime)\right)}\right\}H_1^{(1)}(k\rho).
\nonumber
\eea
with $\Delta\Phi(\varphi,\varphi^\prime)=\frac{\alpha(\varphi^\prime)-\alpha(\varphi)}{2}$ and $\sin\theta_\beta=\frac{\beta}{\sqrt{1+\beta^2}}$. At $\phi=\phi^\prime$, i.e., $\rho=0$, $H_0^{(1)}(k\rho)$ and $H_1^{(1)}(k\rho)$ have a logarithmic and a $1/\rho$ singularity, respectively. In~\eqref{QQB} it is compensated by the prefactor, whereas in~\eqref{QNB} the integral over these singularities leads to the $\sin\theta_\beta$ term on the left hand side of the equation~\cite{Dietz2022}. Accordingly, an interval $[\phi^\prime-\delta\phi,\phi^\prime+\delta\phi]$, where $\delta\phi$ is arbitrarily small, is excluded from the integration range on the right hand side. The corresponding equations for $\psi_2^\ast(\varphi^\prime)$ and $Q_2^{NB}(k;\varphi,\varphi^\prime)$ are obtained with~\eqref{BC1} by multiplying the integrand with $e^{-i\Delta\Phi(\varphi,\varphi^\prime)}$.

For a QB with a mirror symmetry the eigenfunctions can be separated into symmetric and antisymmetric ones with respect to the symmetry axes, and they fulfill either Neumann or Dirichlet BCs along these lines. This is not possible for NBs. Yet, like the eigenfunctions of QBs and GBs, the spinor components of the eigenfunctions of NBs with shapes that exhibit a $Q$-fold rotational symmetry can be separated according to their transformation properties under rotation by $\frac{2\pi}{Q}$ into symmetry classes~\cite{Zhang2021}. The boundary function of billiards with such a shape exhibits a $\frac{2\pi}{Q}$ periodicity,
\bea
w\left(\varphi+\lambda\frac{2\pi}{Q}\right)&=&e^{i\lambda\frac{2\pi}{Q}}w(\varphi),\label{wl}\\
e^{i\alpha\left(\varphi+\lambda\frac{2\pi}{Q}\right)}&=&e^{i\lambda\frac{2\pi}{Q}}e^{i\alpha(\varphi)},\label{alphal}
\eea
with $\lambda =0,1,2,\dots,Q-1$. The eigenstates of the corresponding nonrelativistic QB can be separated into $Q$ subspaces labeled by $l=0,\dots ,Q-1$ according to their transformation properties under the rotation operator $\hat R^{\lambda}$~\cite{Robbins1989,Leyvraz1996}, which produces a rotation by $\frac{2\pi}{Q}$. The eigenfunctions of the QB are characterized by the property
\be
\hat R^{\lambda}\psi^{(l)}_m(r,\varphi)=\psi_m^{(l)}\left(r,\varphi-\frac{2\pi}{Q}\lambda\right)=e^{il\frac{2\pi}{Q}\lambda}\psi^{(l)}_m(r,\varphi).
\label{RotSym}
\ee
This transformation property implies that only the eigenfunctions corresponding to $l=0$ and, for even $Q$ also those with $l=Q/2$, are real and thus invariant under the conventional time-reversal operator $\hat T= \mathcal{\hat C}$ with $\mathcal{\hat C}$ denoting the complex conjugation operator~\cite{Haake2018}. For $l\ne 0,Q/2$ they are complex and
\be
\hat T\psi^{(l)}_m(r,\varphi)=\psi^{(Q-l)}_m(r,\varphi),\label{Top}
\ee
implicating that $\psi_m^{(l)}(r,\varphi)$ and $\psi_m^{(Q-l)}(r,\varphi)$ are eigenfunctions with the same eigenvalue $k_m^2$, because the billiard system is invariant under \T violation. Accordingly, the eigenvalue spectrum of nonrelativistic QBs with a $C_3$ symmetry can be separated into nondegenerate eigenvalues (singlets) with $l=0,\frac{Q}{2}$ and pairwise degenerate ones (doublets) with labels $l,Q-l$. Furthermore, if the corresponding classical dynamics is chaotic and if the billiard boundary has no additional geometric symmetries, the spectral properties of the singlets show GOE behavior, while those of the doublet partners exhibit GUE statistics~\cite{Leyvraz1996}. 

As mentioned above, the spinor components of the eigenstates of the corresponding NB can also be classified according to their transformation properties under a rotation by $\frac{2\pi}{Q}$ into $Q-1$ subspaces~\cite{Leyvraz1996,Keating1997,Robbins1989,Joyner2012}, however, they belong to different ones~\cite{Dietz2021,Zhang2021}. Namely, if the first component of the $m$th spinor eigenfunction belongs to the subspace $l$,
\be
\hat R\psi_{1,m}(\boldsymbol{r})=e^{il\frac{2\pi}{Q}}\psi_{1,m}(\boldsymbol{r}),\label{sympsi1}
\ee
then the second one belongs to the subspace $\tilde l=(l-1)$,
\be
\hat R\psi_{2,m}(\boldsymbol{r})=e^{i(l-1)\frac{2\pi}{Q}}\psi_{2,m}(\boldsymbol{r}),\label{sympsi2}
\ee
where $\tilde l=-1$ corresponds to $l=Q-1$. This intermixture of symmetry classes originates from the additional spin degree of freedom~\cite{Zhang2021,Dietz2021} and is a consequence of the BC~\eqref{BC1}. For all subspaces, the spectral properties of a NB with $Q$-fold symmetry are well described by the GUE, if it has the shape of a billiard with chaotic dynamics and no mirror symmetries. Furthermore, since the Dirac Hamiltonian is not invariant under application of the $\hat T$, the eigenvalues belonging to subspaces $l$ and $Q-l$ are not degenerate. In Refs.~\cite{Zhang2021,Zhang2023c} properties of the eigenstates of GBs with threefold and fourfold symmetry were compared to those of NBs and QBs of corresponding shape, and agreement with those of QBs were found in the nonrelativistic and the relativistic regions. 

\section{Spectral properties of Haldane GBs with three different shapes\label{NR}}

To gain information on universal spectral properties, system-specific ones need to be extracted. This is done by unfolding the eigenvalues to a uniform spectral density, that is, constant mean spacing unity~\cite{Haake2018}. For the unfolding of the eigenvalues of Haldane GBs, we proceeded as in~\cite{Zhang2021} and shifted them such that the smallest eigenvalue equals zero and then replaced the resulting eigenvalues $\tilde E_m$ by the smooth part of the integrated spectral density, $\epsilon_m=N^{smooth}(\tilde E_m)$, which was determined by fitting a second-order polynomial to $N(\tilde E_m)$~\cite{Dietz2015}. Similarly, the ordered eigenwavenumbers $k_m$ of the QBs and NBs were unfolded to mean spacing unity by replacing them by the smooth part of the integrated spectral density, $\epsilon_m=N^{smooth}(k_m)$, which is given by Weyl's formula~\cite{Weyl1912}, $N^{Weyl}(k_m)=\frac{\mathcal{A}}{4\pi}k_m^2-\frac{\mathcal{L}}{4\pi}k_m+C_0$, with $\mathcal{A}$ denoting the area. For massless NBs the perimeter term is absent~\cite{Berry1987}. We present results for the Dyson-Mehta statistics, $\Delta_3(L)$, of the spectrum~\cite{Mehta1990}, which is defined as the least-squares deviation of the integrated spectral density of the unfolded eigenvalues from the straight line best fitting it in the interval $L$ and provides a measure for the degree of rigidity of a level sequence. Furthermore, we analyzed the distribution of the ratios~\cite{Oganesyan2007,Atas2013} of consecutive spacings between nearest neighbors, $r_m=\frac{\epsilon_{m+1}-\epsilon_{m}}{\epsilon_{m}-\epsilon_{m-1}}$, which are dimensionless so that no unfolding is needed~\cite{Oganesyan2007,Atas2013,Atas2013a}. Analytical results have been obtained for the average ratios, $\langle r\rangle$ and also for $\tilde r_m={\rm min}\left(r_m,\frac{1}{r_m}\right)$ in Ref.~\cite{Atas2013a}, $\langle r\rangle\simeq 1.75,1.36$ and  $\langle\tilde r\rangle \simeq 0.54,0.60$ for the GOE and GUE, respectively, and for Poissonian random numbers $\langle r\rangle =\infty$ and $\langle\tilde r\rangle \simeq 0.39$.

We analyzed the spectral properties of the Haldane GBs around the band edges and the Dirac point. The DOS is symmetric with respect to the Dirac point. Therefore, we restricted to the eigenvalues at the lower band edge and above the Dirac point, respectively. Here, we excluded the edge states, that are present for the case $M=0$ around $E=0$, and lead to an exceptionally high DOS around that energy value~\cite{Wurm2009,Wurm2011}. Their contributions are non-universal due to the localization properties of the associated wave functions, implicating deviations from random-matrix theory (RMT) predictions~\cite{Dietz2015,Dietz2016} for GBs with the shape of a chaotic CB. 

\subsection{Haldane GBs with rectangular shape}

Rectangular Haldane billiards have two mirror symmetries and a twofold rotational symmetry. Accordingly, the eigenfunctions of the QB and GB, and the spinor components of the eigenfunctions of the NB can be classidieded according to their transformation properties under rotation by $\frac{\pi}{2}$~\cite{Zhang2021}. We exploited this property in~\cite{Dietz2023} and found out that the spectral properties of the symmetry-projected eigenstates of rectangular NBs corresponding to either of the two symmetry classes exhibit semi-Poisson statistics, whereas they agree with Poisson statistics when we consider the whole spectrum irrespective of the symmetries. Here, a sequence of random numbers with semi-Poisson statistics is obtained from one with Poisson statistics by sorting the numbers by size and deleting every second one. The nearest-neighbor spacing distribution of the eigenvalues of rectangular QBs, whose ratio of side lengths is a rational number, exhibit gaps, that is, they are untypical integrable systems. In order to realize a rectangular QB whose eigenvalue spectrum exhibits short-range correlations that comply with Poisson statistics, the ratio of side lengths needs to be an irrational number~\cite{Elkies2004}. The long-range correlations, on the other hand, approach Poisson statistics with an increasing number of eigenvalues for rational and irrational ratios~\cite{Marklof1998,ElBaz2015}. 
\begin{figure}
        \begin{center}
        \includegraphics[width=0.8\textwidth]{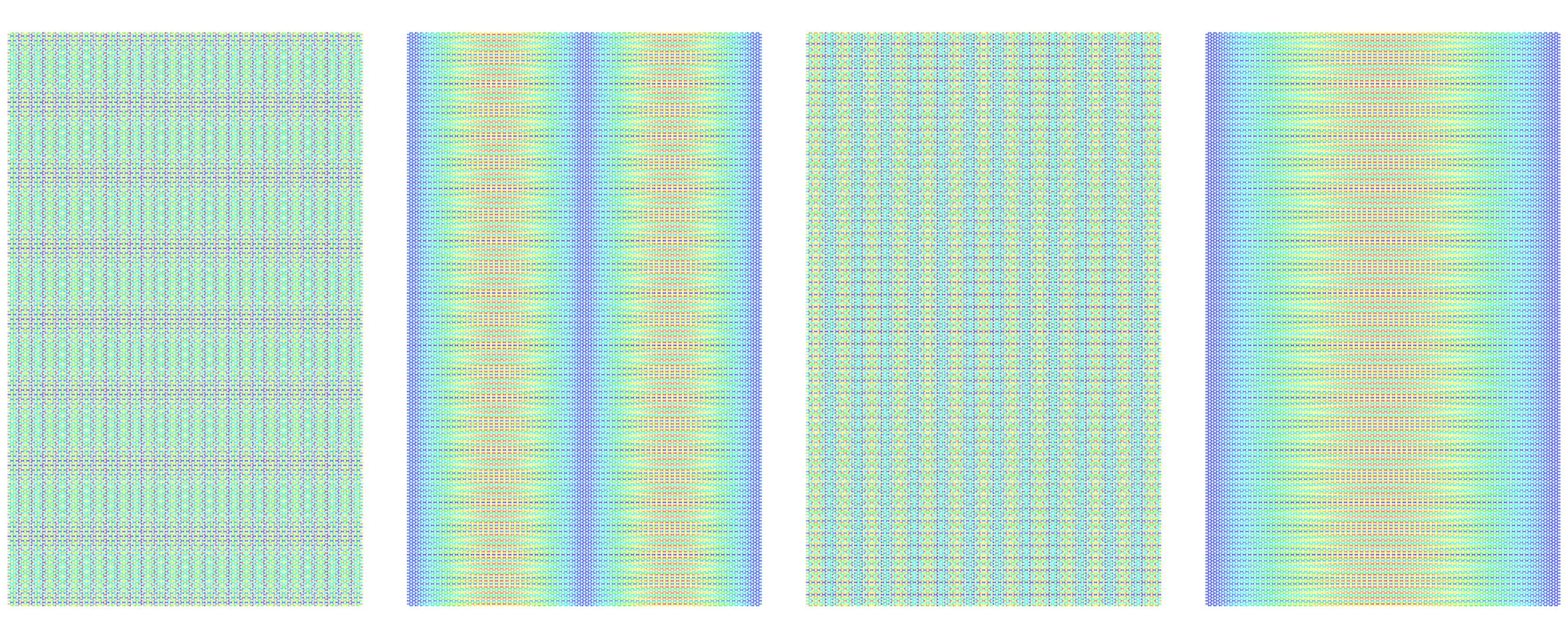}
\caption{Examples for wave functions of the rectangular GB (Haldane GB with mass $M=0.0$) in the region $E\gtrsim -0.5$ below the Dirac point.}
        \label{fig:WFsRectGB0}
\end{center}
\end{figure}
\begin{figure}
	\begin{center}
	\includegraphics[width=0.8\textwidth]{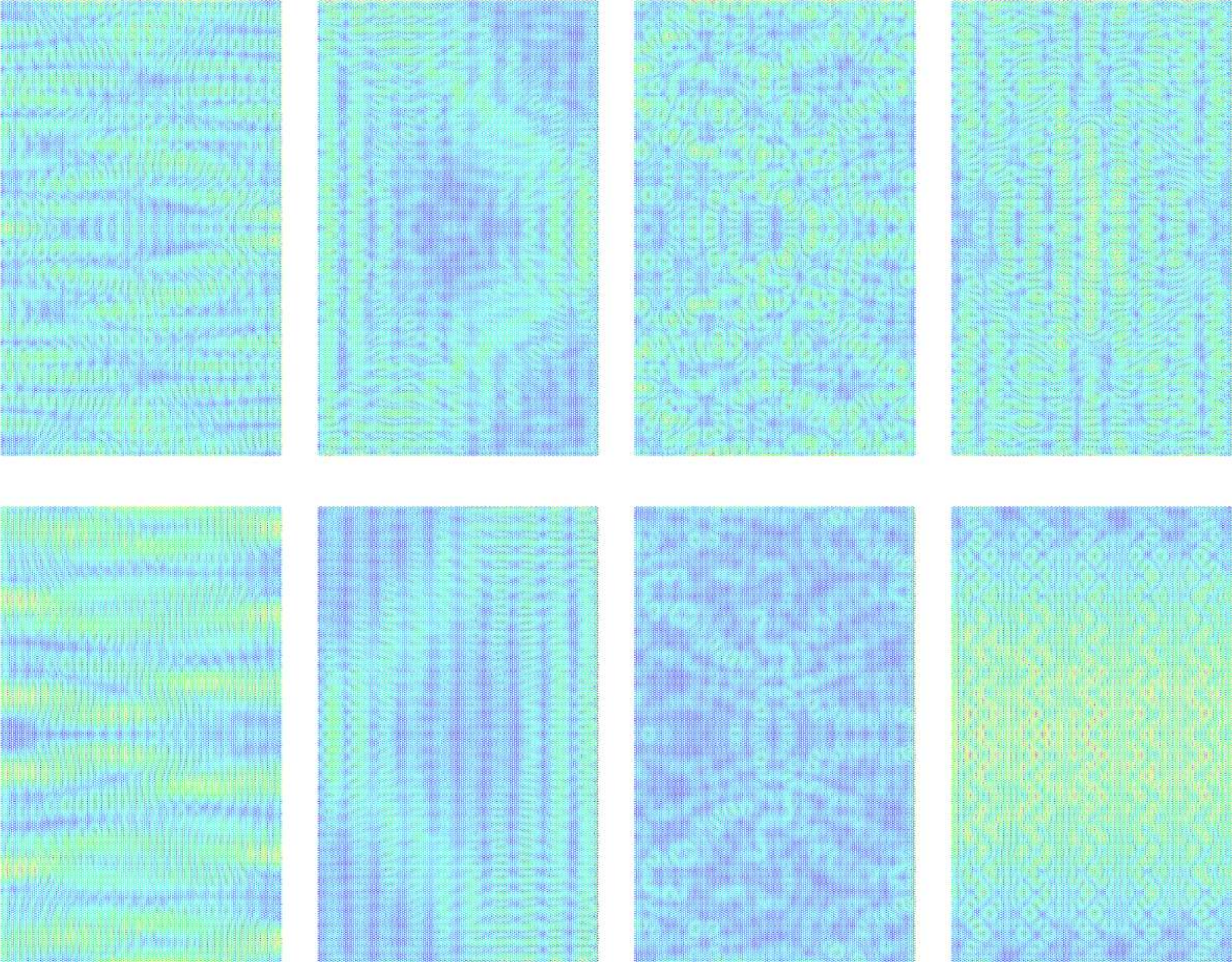}
\caption{Examples for wave intensity distribution of the rectangular Haldane GB with mass $M=0.3$ in the region $E\gtrsim -0.5$ below the Dirac point, where the band structure is gapped at the $K^\prime$ point.}
	\label{fig:WFsRectGB}
\end{center}
\end{figure}
\begin{figure}
	\begin{center}
    \includegraphics[width=0.625\textwidth,angle=90]{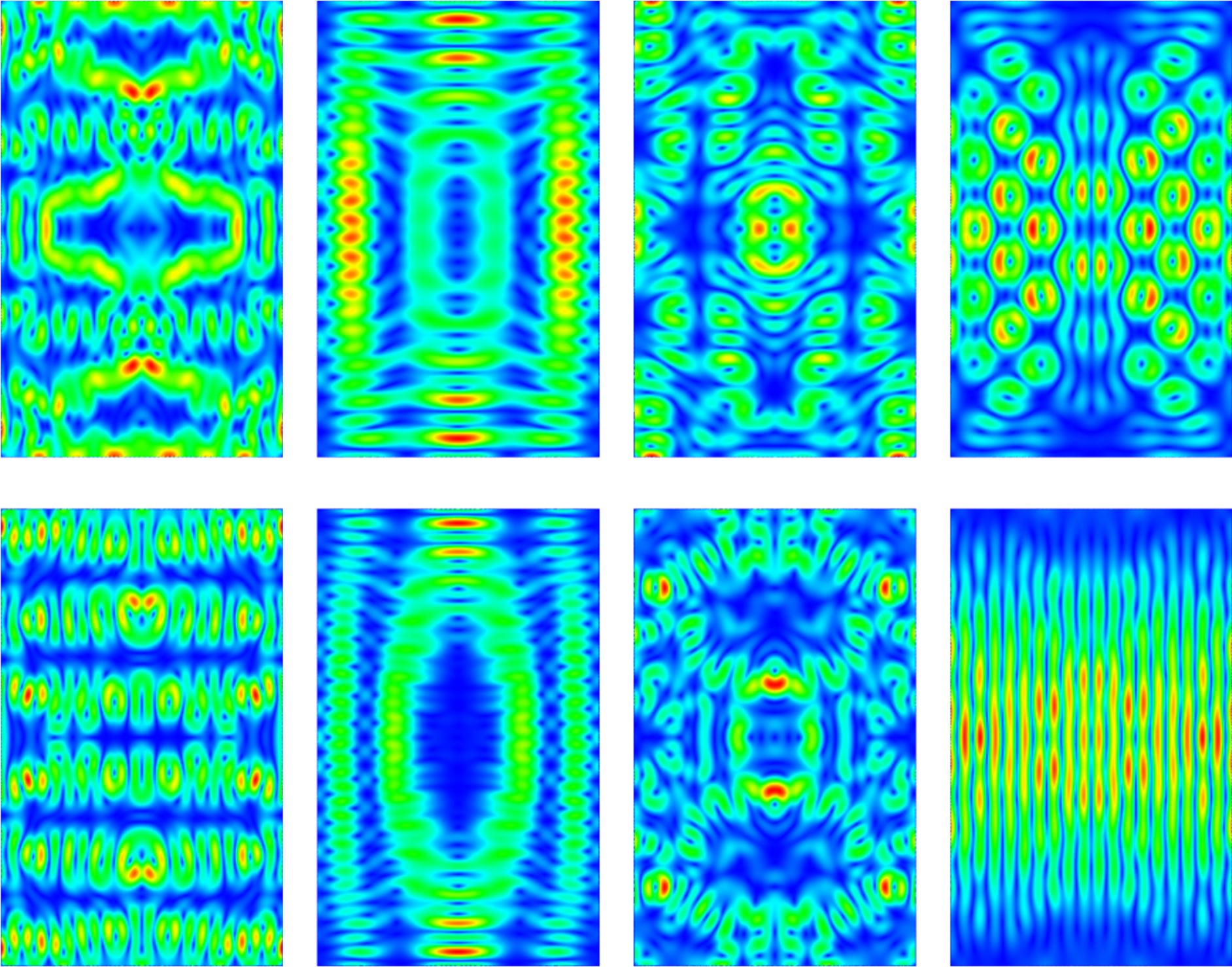}
\caption{Modulus of the local current for six spinor eigenfunctions of the rectangular NB with mass $M=0.3$. In the first row examples are shown for which the first spinor component is symmetric, the second one is antisymmetric with respect to rotation by $\pi$, in the second one the first one is antisymmetric and the second one is symmetric.}
	\label{fig:WFsRectNB}
\end{center}
\end{figure}
\begin{figure}
	\begin{center}
	\includegraphics[width=0.7\textwidth]{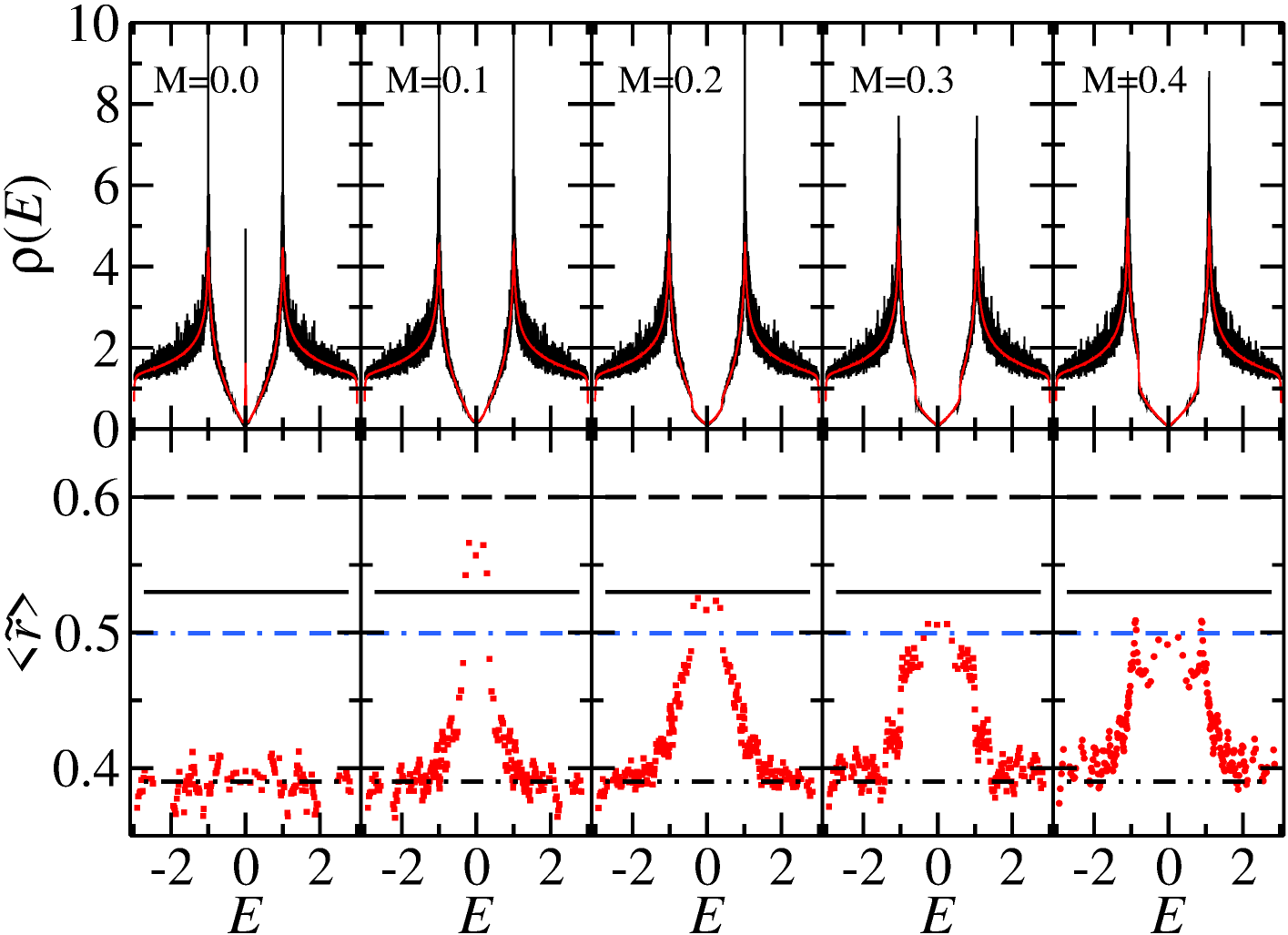}
\caption{Upper row: DOS of the rectangular Haldane GB with ratio of side lengths equal to the golden mean. Lower row: Corresponding average ratios $\langle\tilde r\rangle$. The black dashed, solid, and dash-dotted lines mark the values for GUE, GOE and Poisson, respectively, and the blue dash-dash-dotted line that for semi-Poisson. For $M=0.0$ they agree with Poisson, and for $M\gtrsim 0.2-0.4$ they are close to semi-Poisson around the Dirac point.}
	\label{fig:RhoRect}
\end{center}
\end{figure}
\begin{figure}
	\begin{center}
	\includegraphics[width=0.7\textwidth]{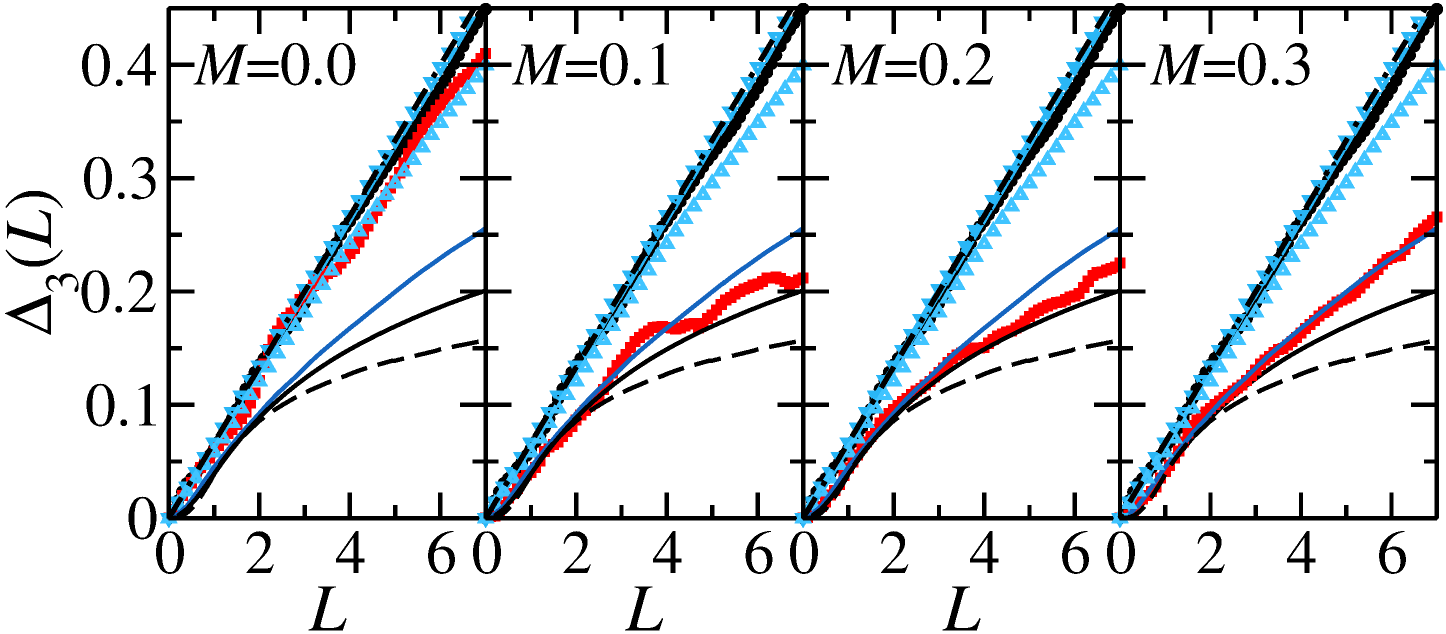}
	\caption{Dyson-Mehta statistics for the rectangular Haldane GB with ratio of side lengths equal to the golden mean around the band edge (black dots) and around the Dirac point (red squares). The light-blue triangles down and up curves show $\Delta_3(L)$ for the quantum and neutrino billiard, respectively, with same ratio of side lengths $\frac{L_y}{L_x}=\frac{1+\sqrt{5}}{2}$. The number of eigenvalues is only 100 around the Dirac point for $M=0.1$, whereas for the other cases it equals 200. We obtain the interesting result for the energy spectrum around the Dirac point, that with increasing mass a transition from Poisson to semi-Poisson takes place. The black dashed, solid, and dash-dotted lines mark the values for GUE, GOE and Poisson, respectively. The blue curve shows the result for the symmetry-projected states of the rectangular NB with $l=0$, which is close to that for semi-Poisson statistics (see main text).}
	\label{fig:D3Rect}
\end{center}
\end{figure}

The ratio of the side lengths $L_y$ and $L_x$ was chosen equal to the golden mean, $\frac{L_y}{L_x}=\frac{1+\sqrt{5}}{2}$, and $L_x$ was chosen such that a honeycomb lattice with 49608 sites fits into the billiard domain. Figure~\ref{fig:WFsRectGB0} depicts four examples of wave functions of the Haldane GB for $M=0.0$ around $E\gtrsim -0.5$ below the Dirac point. The wave functions exhibit patterns that are typical for the QB of corresponding shape. This changes drastically when turning on the mass term, $M\gtrsim 0$. In~\reffig{fig:WFsRectGB} we show for $M=0.3$ eight examples for wave functions around $E\simeq -0.5$, that is, close to the lower critical value, $E\gtrsim -2M=-0.6$, with one Dirac cone at the $K$ point and no excitations at the $K^\prime$ point. In the first and second rows of the first three columns we show wave functions that seem to be symmetry related, and in the last column and bottom row one example for a trivial eigenmode, that bounces back and forth between the two longer sides. In~\reffig{fig:WFsRectNB} examples of the modulus of the local current defined in~\eqref{eq:curr} across the billiard area are plotted for the corresponding NB, that exhibit similar pattern structures as the wave functions in~\reffig{fig:WFsRectGB}.  In the upper row we show examples for which the first spinor component is symmetric and the second one is antisymmetric under rotation by $\pi$, whereas in the second one it is antisymmetric for the first spinor component and symmetric for the second one. Since the local current depends on products of the first and second one, it is in all cases antisymmetric under rotation by $\pi$ as confirmed by its phase distribution (not shown). An observation is warranted here. For the Haldane GB, we encounter a loss of $C_2$ symmetry alongside a disruption of the valley symmetry, evident in the pattern of the wave functions. 

The upper row of~\reffig{fig:RhoRect} exhibits the DOS for $M=0.1-0.4$. The peak at the Dirac point, visible at $E=0$ for $M=0$, results from the edge states that are localized at the zigzag edges. At $\vert E\vert =2M$ the DOS exhibits a jump. Below that value, conical band touching is only present at the $K$ point, as in the case illustrated in~\reffig{fig:Bandst} for a Haldane GB with the shape of a Africa billiard. Indeed, this feature is observed, independently for all considered shapes, as outlined below. The lower row shows the corresponding average ratios $\langle\tilde r\rangle$.  The dashed, solid and dash-dotted lines mark the values for GUE, GOE and Poisson, respectively. For $M=0$ they agree with that of Possonian random numbers, whereas, when increasing $E$ starting from the lower band edge or decreasing it starting from the upper one $\langle\tilde r\rangle$ takes a value intermediate between that for semi-Poisson statistics (blue dash-dash-dotted line) GOE statistics. For $M=0.2-0.4$ it approaches the semi-Poisson value in the region with $\vert E\vert\leq 2M$. 

Similar results are obtained for other statistical measures of short- and long-range correlations in the eigenvalue spectra. In~\reffig{fig:D3Rect} we show the results for the Dyson-Mehta statistics. Here, the squences comprised 500 eigenvalues around the band edge and in the region of linear dispersion around the $K$ point for $M=0$ they contained 200 levels, also for $M=0.2-0.4$. However, for $M=0.1$ there are only 100 eigenvalues in the region of linear dispersion for $\vert E\vert\leq 2M$. Accordingly, around the $K$ point the statistics is worse than around the band edges. Yet, with increasing $M$ the curve approaches the blue one showing the result for the symmetry-projected eigenstates of the corresponding NB, which exhibits semi-Poisson statistics~\cite{Dietz2023}. Indeed, introducing the mass $M$ and the \Ti-violating term, implicating the disappearance of the Dirac cone at either the $K$ or the $K^\prime$ point, corresponds to a symmetry-projection onto one of the pseudo-spins.

\subsection{Haldane GBs with the shape of an Africa billiard}

Africa-shaped billiards provide a paradigm system for the study of the spectral properties of fully chaotic systems, since their boundary does not comprise regions where bouncing-ball orbits~\cite{Sieber1993} may exist that bounce back and forth between opposite sides or orbits that are confined to a fractional part of the available phase space. Such orbits do not feel the chaoticity of the dynamics generated by the boundary and lead to scarred wave functions in the corresponding quantum system and, therefore, to deviations of the spectral properties from RMT predicitions~\cite{Heller1984} of the QB, also of the NB~\cite{Zhang2021} of corresponding shape. The domain of the Africa billiard~\cite{Berry1986,Berry1987} is defined by the parameterization
\be
x(r,\phi)+iy(r,\phi)=w(r,\phi)=r\left[\zeta+0.2\zeta^2+0.2\zeta^3e^{i\frac{\pi}{3}}\right], \phi\in [0,2\pi),\, r\in[0,r_0],
\label{Africa}
\ee
where we introduced the notation $\zeta(\phi)=e^{i\phi}$. Here, $r_0$ was chosen such that a honeycomb lattice with 65199 sites fits into the billiard domain.

\begin{figure}
	\begin{center}
	\includegraphics[width=0.7\textwidth]{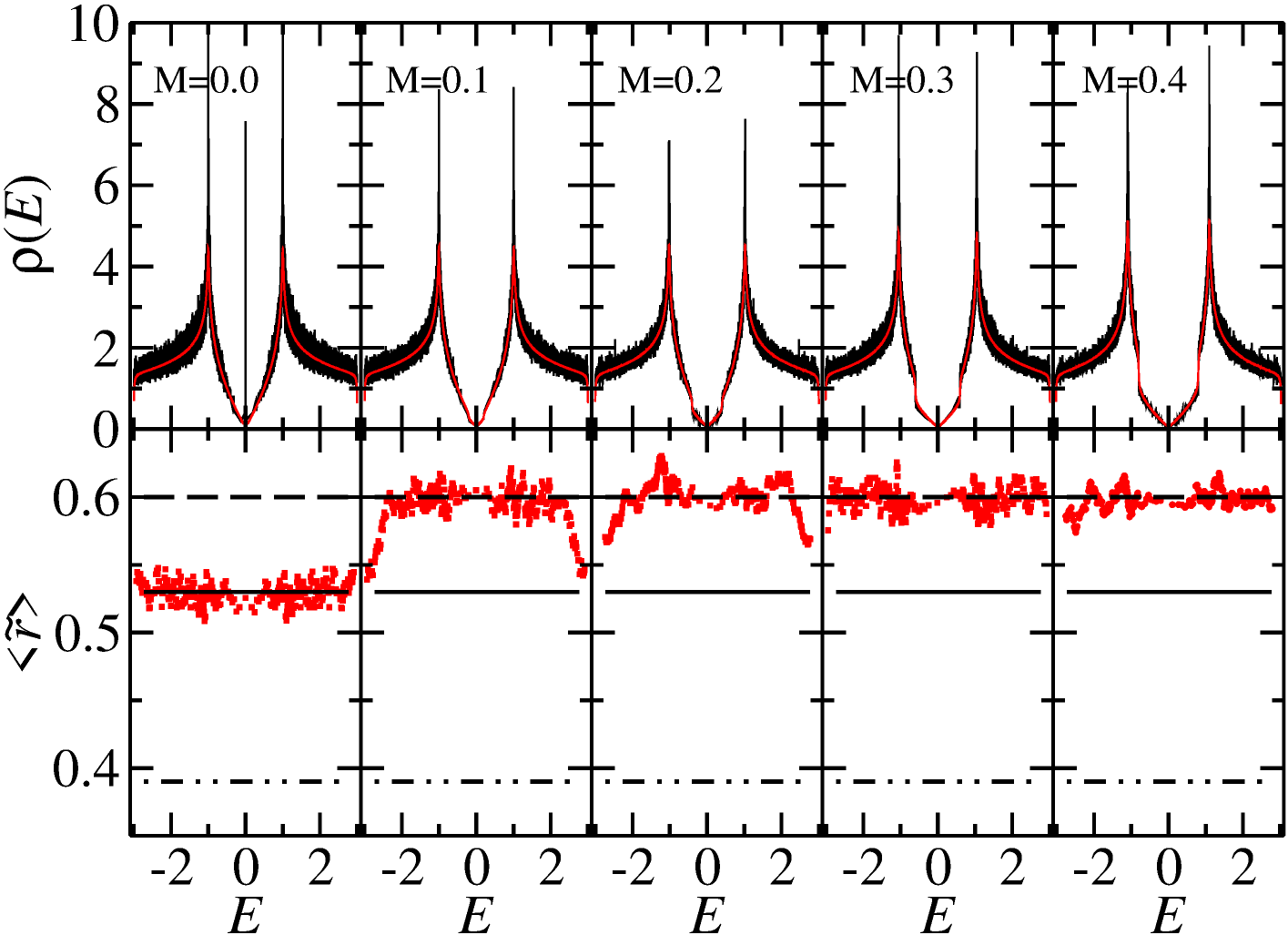}	
\caption{Same as~\reffig{fig:RhoRect} for the Africa shape. For $M=0.0$ the average ratios agree with GOE, for $M\gtrsim 0.1$ with GUE for $\vert E\vert\lesssim 2$, with GOE around the band edges at $\vert E\vert\simeq 3$, whereas their statistics is between GOE and GUE otherwise.}
	\label{fig:RhoAF}
\end{center}
\end{figure}

\begin{figure}
	\begin{center}
	\includegraphics[width=0.7\textwidth]{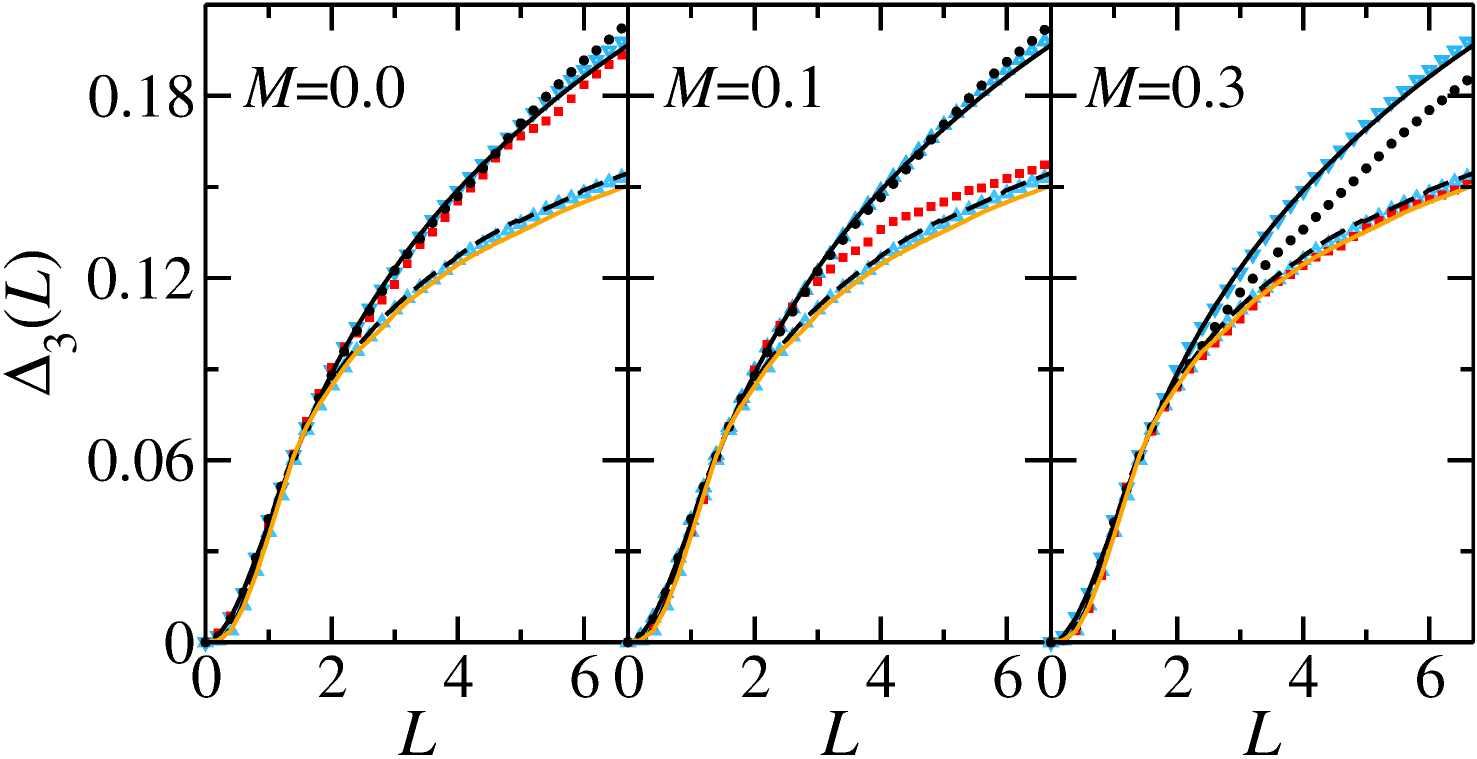}	
\caption{Same as~\reffig{fig:D3Rect} for the Africa shape. It agrees with GOE for all three masses around the band edge. Around the Dirac point the $\Delta_3$-statistics agrees with GOE for $M=0.0$, for $M=0.1$ it is between GOE and GUE, and for $M=0.3$ it agrees with GUE. The orange curve shows an example for the massive NB with mass $k\beta =3$. It is close to the GUE curve and also agrees well with the result for the Haldane GB.} 
	\label{fig:D3AF}
\end{center}
\end{figure}

The upper row of~\reffig{fig:RhoAF} shows the DOS of the Haldane GB with that shape, the lower one the average ratios $\langle\tilde r\rangle$. For $M=0$ the DOS exhibits a peak at $E=0$ which is due to edge states at zigzag edges of the Haldane GB~\cite{Wurm2011}. Again, a jump is visible at $E=\pm 2M$ for all masses. The average ratios fluctuate about the value for the GOE for $M=0$, also for $\vert E\vert\gtrsim 2$ and around the GUE otherwise for $M\gtrsim 0.1$. 
Figure~\ref{fig:D3AF} exhibits the Dyson-Mehta statstics $\Delta_3(L)$ for the Haldane GB with the shape of a Africa billiard for $M=0,0.1,0.3$ around the band edges (black dots) and around the Dirac point (red squares). Furthermore, the results for the QB (light-blue down triangles) and the NB (light blue up triangles) are plotted. Around the Dirac point and for $M\gtrsim 0.1$ it agrees well with the GUE curve. In~\cite{Zhang2021,Zhang2023c} we also computed the eigenstates for massive NBs and showed that agreement with the spectral properties of GBs of corresponding shape are only found for large masses $k\beta\gtrsim 100$, that is, in the nonrelativistic limit. In~\reffig{fig:D3AF} we include the results for a small mass $k\beta =3$, for which we find good agreement with those of the massless NB and also the Haldane GB.   

\subsection{Haldane GBs whose shape has a $C_3$ symmetry}
The domain of the Haldane GB with $C_3$ symmetry, referred to as '$C_3$ Haldane GB' in the following, is defined in the $(x,y)$ plane by the parametrization
\bea
&&x(r,\phi)+iy(r,\phi)=w(r,\phi)=R(r,\phi)e^{i\phi},\\
&&\phi\in [0,2\pi),\, r\in [0,r_0],\nonumber
\label{coordinates}
\eea
with
\bea
&&R(r,\phi)=rf(\phi),\\
&&f(\phi)=1+0.2\cos(3\phi)-0.2\sin(6\phi).
\eea
Here, $r_0$ was chosen such that a honeycomb lattice with $3\times 24189$ sites fits into the billiard domain.
 
\begin{figure}
        \begin{center}
    \includegraphics[width=0.65\textwidth]{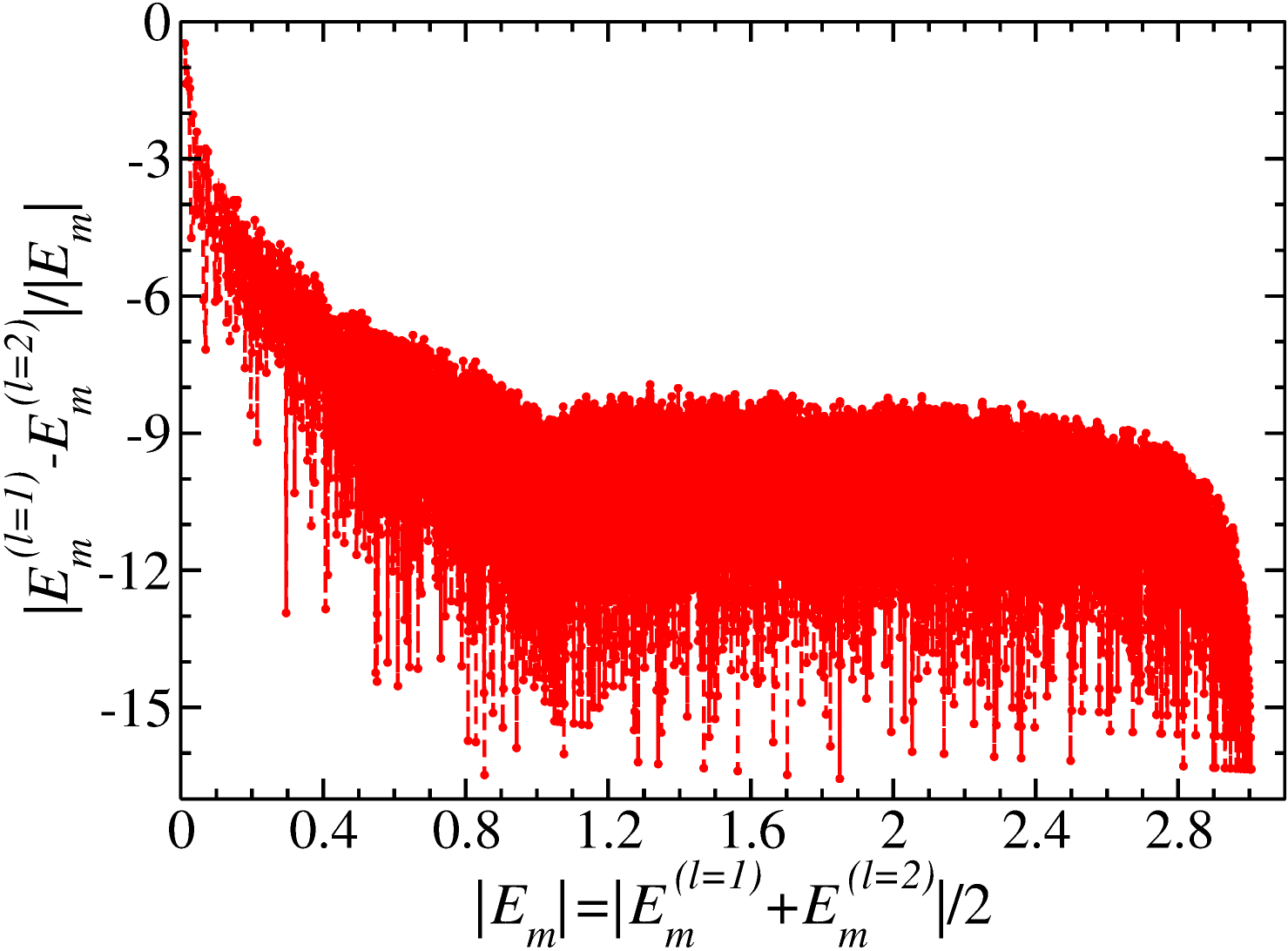}
\caption{Relative distances $\frac{\vert E^{(l=1)}_m-E^{(l=2)}_m\vert}{\vert E^{(l=1)}_m+E^{(l=2)}_m\vert}$ between corresponding eigenvalues of the $C_3$ Haldane GB from the symmetry-projected subspectra with $l=1$ and $l=2$, respectively, for $M=0.2$. In the nonrelativistic region around the band edges at $\vert E\vert = 3$ they are degenerate. In contrast, in the relativistic region around the $K$ point at $E=0$ the degeneracy is clearly lifted as expected for systems governed by the Dirac equation~\eqref{DE}, like for NBs (see~\refsec{NM}).}
\label{fig:DiffC3}
\end{center}
\end{figure}

\begin{figure}
	\begin{center}	
    \includegraphics[width=0.65\textwidth]{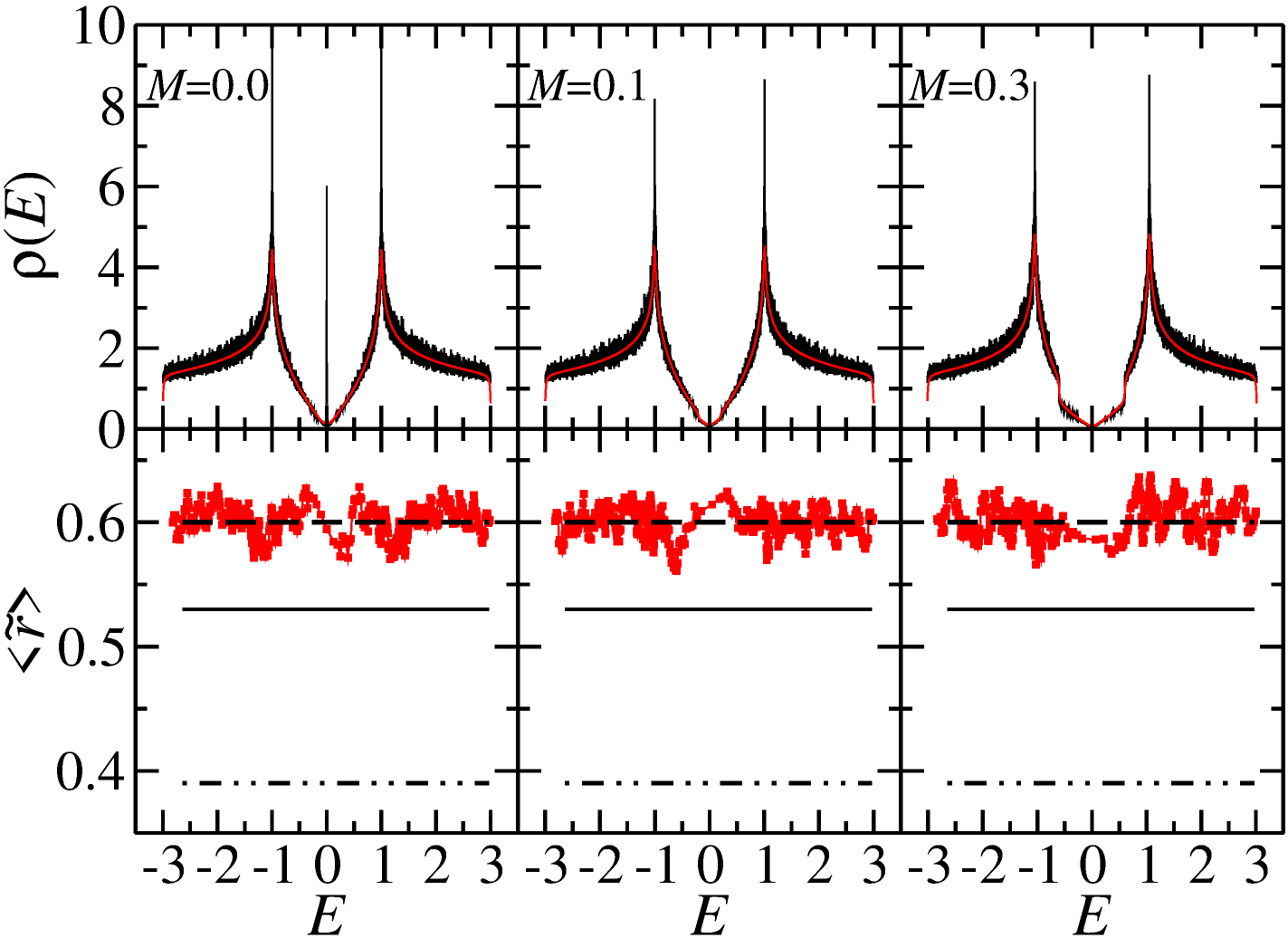}	
\caption{Same as~\reffig{fig:RhoAF} for the $C_3$ Haldane GB for the symmetry class $l=1$. The average ratios agree with GUE for all masses. }
	\label{fig:RhoC3}
\end{center}
\end{figure}

\begin{figure}
	\begin{center}
	\includegraphics[width=0.65\textwidth]{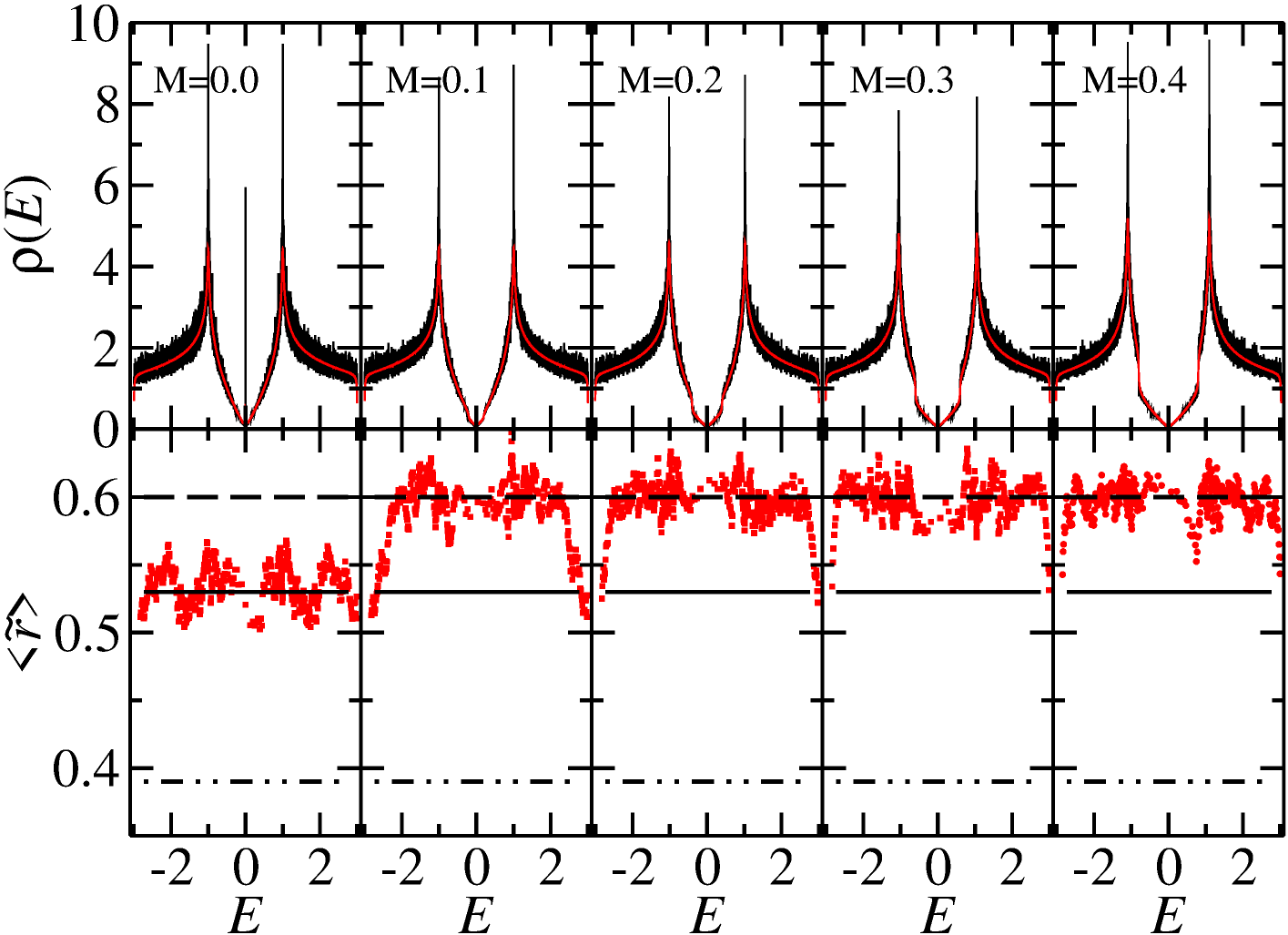}	
\caption{Same as~\reffig{fig:RhoC3} for the $C_3$ Haldane GB for the symmetry class $l=0$. For $M=0.0$ the average ratios agree with GOE, for $M=0.1$ with GOE below $\vert E\vert\gtrsim 2.5$, and with GUE otherwise, for $M\gtrsim 0.2$ they agree with GUE.}
	\label{fig:RhoC3_f}
\end{center}
\end{figure}

\begin{figure}
	\begin{center}
    \includegraphics[width=0.7\textwidth]{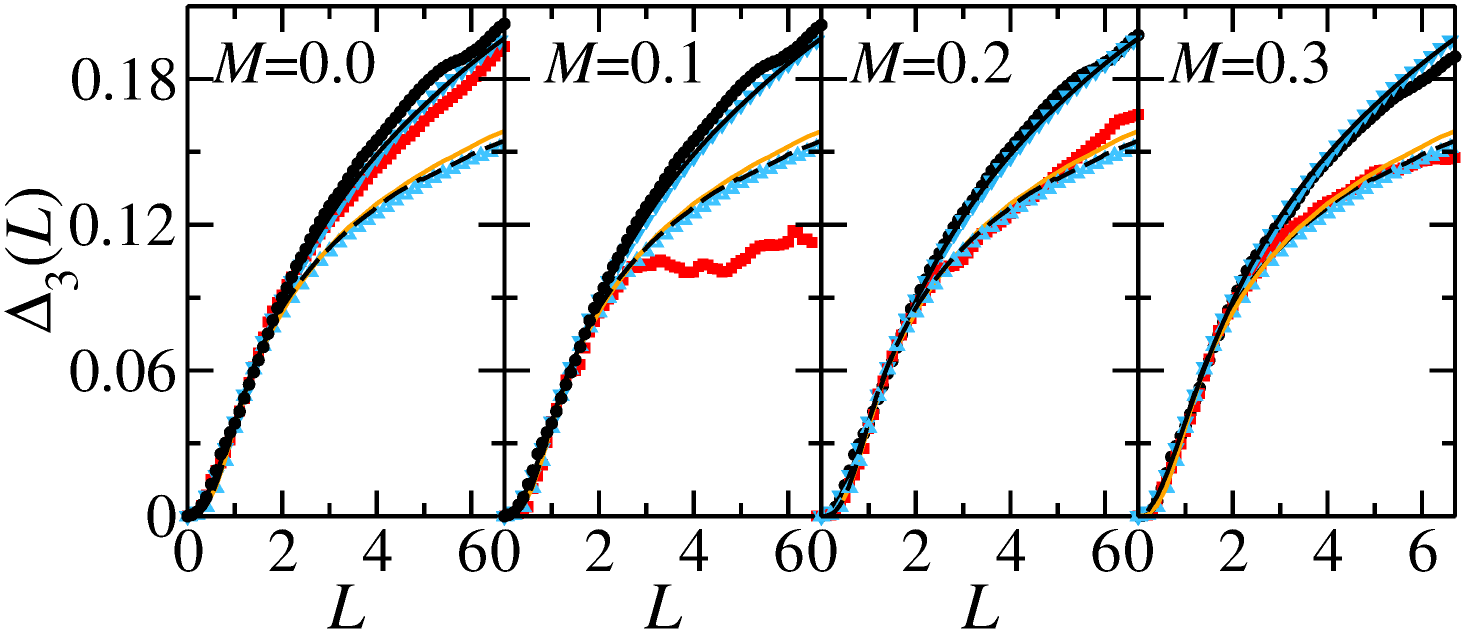}
\caption{Same as~\reffig{fig:D3AF} for the $C_3$ Haldane GB for the symmetry class $l=0$. It agrees with GOE for all three masses around the band edge. Around the Dirac point the $\Delta_3$-statistics agrees with GOE for $M=0.0$, for $M=0.1$ it is close to GUE, and saturates at $L\simeq 2$, the reason being that there we have only 100 eigenvalues, and for $M=0.3$ it agrees with GUE.}
	\label{fig:D3C3}
\end{center}
\end{figure}

Due to the $C_3$ rotational symmetry, the eigenstates of the QB can be separated into three symmetry classes labeled by $l=0,1,2$. For $l=0$ the eigenfunctions are invariant under rotation by $\frac{2\pi}{3}$ and for $l=1,2$ the eigenstates are turned into each other when applying $\hat T$. Accordingly, the eigenvalue spectrum can be separated into singlets with $l=0$ and pairs of degenerate eigenvalues corresponding to $l=1,2$, that exhibit GOE and GUE statistics, respectively. Similarly, the spinor components can be separated into three symmetry classes labeled by $l=0,1,2$. Here, for $l=0$ the first spinor component of $\boldsymbol{\psi}$ is invariant under rotation by $\frac{2\pi}{3}$. Since \Ti invariance is violated, the symmetry-projected spectra corresponding to $l=1$ and $l=2$ are not degenerate and the spectral properties agree with GUE statistics for all three symmetry classes. Figure~\ref{fig:DiffC3} exhibits the difference between corresponding eigenvalues for $l=1$ and $l=2$ for mass $M=0.2$. For $\vert E\vert\gtrsim 1.0$ the eigenvalues are degenerate as in the case of the nondegenerate QB. However, in the relativistic region $\vert E\vert\lesssim 1$ the degeneracy is clearly lifted as for the NB of corresponding shape. In~\reffig{fig:RhoC3} we show the DOS and ratios for $l=1$. The DOS exhibits the same features as for the Africa-shaped Haldane GB (see~\reffig{fig:RhoAF}) whereas the ratios coincide with GUE for all masses. On the contrary, as illustrated in~\reffig{fig:RhoC3_f}, for $l=0$ the ratios are close to the GOE for $M=0.0$ and close to the GUE for $\vert E\vert\lesssim 2$ and $M\gtrsim 0.1$. In ~\reffig{fig:D3C3} we compare the spectral properties of the Haldane GB around the band edges (black dots) and Dirac point (red squares) with those of the nonrelativistic QB (light-blue down triangle), the massless NB (light-blue up triangles) and for a small mass $k\beta =5$ (orange solid line). For $M\gtrsim 0.1$ the curve for the Haldane GB is closer to that for the massless NB  than to that for the massive one. Thus we again find good agreement between the spectral properties for the Haldane GB with $M\gtrsim 0.1$ and the massless NB. 

\section{Conclusions\label{Concl}}

In summary, we introduced the honeycomb-lattice based Haldane model as an alternative to GBs for the simulation of universal features of the eigenstates of relativistic neutrino billiards. The Haldane model offers a more suitable framework as it suppresses inter-valley scattering inherent in the finite-size  artificial-graphene model. We explore numerically the critical Haldane model behavior on rectangular, Africa, and $C_3$ shapes. These simulations demonstrate a phase transition from non-relativistic to relativistic quantum behavior upon adjusting the Haldane tunneling parameters, thereby affirming the efficacy of the critical Haldane model in mimicking relativistic neutrino phenomena within a tight-binding framework. With the recent proposals to simulate the Haldane model with photonic crystals~\cite{HDphotonic1,HDphotonic2}, we expect the feasibility of generating energy spectra exhibiting the particular phenomena of NBs experimentally. 

In forthcoming research, we aim to investigate the non-critical Haldane model precisely when $t_2 \neq \frac{M}{3\sqrt{3}}$, offering the potential to simulate neutrino dynamics incorporating the mass term that can be finely tuned employing the control parameters $t_2$ and $M$ of the Haldane model. Additionally, we plan to delve deeper into the tight-binding model of A-B stacking bilayer graphene, anticipating quadratic band intersections at low energy limits~\cite{McCann2013},  thereby enabling simulations of a new type of quantum billiards.

\section{Acknowledgement} 
We acknowledge financial support from the Institute for Basic Science in Korea through the project IBS-R024-D1.

\bibliography{BD_Haldane}
\end{document}